\documentclass[prx, reprint,
superscriptaddress,
 amsmath,amssymb,
 aps,
floatfix,
longbibliography,
]{revtex4-2}

\usepackage{graphicx}
\usepackage{dcolumn}
\usepackage{bm}
\usepackage{amsmath}

\begin{document}

\title{Lossless monochromation for electron microscopy \\ with pulsed photoemission sources and rf cavities}%

\author{C. J. R. Duncan}%

\affiliation{Cornell Laboratory for Accelerator-Based Sciences and Education, Cornell University,
Ithaca, New York 14853, USA
}
\author{D. A. Muller}
\affiliation{School of Applied and Engineering Physics, Cornell University, Ithaca, New York 14853 USA}
\author{J. M. Maxson$^1$}
\email{jmm586@cornell.edu}

\begin{abstract}

   Resonant radiofrequency cavities enable exquisite time-energy control of electron beams when synchronized with laser driven photoemission. We present a lossless monochromator design that exploits this fine control in the one-electron-per-pulse regime.  The theoretically achievable maximum beam current on target is orders of magnitude greater than state-of-the-art monochromators for the same space-time-energy resolution. This improvement is the result of monochromating in the time domain, unconstrained by the transverse brightness of the electron source.  We show analytically and confirm numerically that cavity parameters chosen to minimize energy spread perform the additional function of undoing the appreciable effect of chromatic aberration in the upstream optics. We argue that our design has significant applications in ultrafast microscopy, as well as potential for use in non-time resolved microscopy, provided future photoelectron sources of sufficiently small size and laser sources of sufficiently high repetition rate. Our design achieves in simulations more than two orders of magnitude reduction in beam energy spread, down to single digit meV. Overcoming the minimum probe-size limit that chromatic aberration imposes, our design clears a path for high-current, high-resolution electron beam applications at primary energies from single to hundreds of keV.

\end{abstract}

\maketitle

\section{Introduction}

The electron-optical properties of time varying fields have long been of fundamental interest to electron microscopists and accelerator physicists \cite{Hawkes2009a, scherzer1949theoretical, schonhense2002correction, Khursheed2005, Reijnders2010, oldfield1974microwave}. Resonant radio-frequency (rf) cavities in particular have found significant use over the past two decades in time-resolved electron diffraction and microscopy. Highlights of a versatile range of applications include: compressing electron pulses in time to the femtosecond scale and below \cite{Chatelain2012a, Gliserin2015,VanOudheusden2010, Maxson2017, zeitler2015linearization, pasmans2013microwave, Zhao2018, Li2019}, temporal magnification of ultrafast events \cite{Cesar2019, Kolner1990}, impulsive acceleration and deceleration of beams over short distances \cite{Weathersby2015, Zhao2018, Hachmann2016,  Filippetto2016, Sannomiya2019}, chopping continuous beams into short pulses \cite{Sannomiya2019, Lassise2012, Oldfield1976, Verhoeven2018, murooka2011transmission},  and the controlled introduction of time-of-arrival correlations for performing energy measurements \cite{Verhoeven, Zhao2018}. Radio-frequency cavities are the workhorses of particle acceleration at primary energies above $1 \ \mathrm{MeV}$. Efforts to adopt rf technology for sub-MeV electron microscopes have faced the demanding requirement for fine precision in the timing of electron bunches. Early work on beam chopping at the picosecond scale resulted in the development of a GHz repetition rate SEM \cite{ura1973generation, hosokawa1978gigahertz}, but contemporaneous investigation showed that picosecond precision was insufficient to realize the theoretical potential of cavities as high-quality electron lenses \cite{oldfield1974microwave}.  Today, advances in broadband laser sources and photocathode materials have allowed synchronising electron pulses with rf phases to single femtoscond precision \cite{otto2017solving}, setting the stage for a new round in the fruitful exchange of expertise between the accelerator and microscopy research communities.

The focus of the present work is the use of cavity fields to compress beam energy spreads in the single-electron-per-pulse regime. Lower energy spread beams are advantageous in all electron microscopy, both static and time resolved. The importance to electron energy loss spectroscopy (EELS) is clear: monochromation of continuous-wave (CW) cold field emission (CFE) sources to the single meV scale has enabled measurements of phonon spectra  with atomic spatial resolution in the scanning transmission electron microscope \cite{Krivanek2014, venkatraman2019vibrational}. Ultrafast pulsed beams of equally narrow energy spread would make it possible to resolve the different contributions of coupled lattice, charge, and spin dynamics to the spectrum of quantum materials \cite{carbonereview} . In applications beyond spectroscopy, source energy spread limits the spatial resolution of electron microscopes. Chromatic aberration is the barrier to achieving atomic diameter probes at low primary energies of less than $5 \ \mathrm{keV}$ \cite{joy1996low}, a commonplace regime for scanning electron microscopy and industrial meteorology methods, such as time-resolved cathodoluminescence. Finally, designs for damage-mitigating, pulsed multipass electron microscopes employ monochromation \cite{Koppell2019}. 

The challenge in producing low-energy spread beams is that the best sources have intrinsic spreads of hundreds of meV.  To date, monochromator designs that reach the single meV energy scale have relied on apertures in energy-dispersive locations \cite{krivanek2009high, Mankos2016}. Apertures cause a loss of beam current by a factor equal to the ratio between the desired energy spread and the source energy spread.  A factor 100 reduction leaves little current for imaging in the continuous case \cite{Krivanek2014}, and prohibitively low current in the ultrafast case. Lossy monochromation at low voltages is infeasible because beam current must increase as accelerating voltage decreases to maintain a tolerable detector signal.

Pulsed sources combined with rf fields provide a direct experimental handle on the beam's longitudinal phase space, comprising the conjugate dynamical variables of forward momentum and time of arrival at a given transverse plane \cite{Williams2017, pasmans2013microwave}.  Photoemission is capable of delivering subpicosecond electron pulses with femtosecond timing precision to experimental targets. Lossless energy spread reduction is therefore possible because the time of arrival --- and hence rf accelerating phase --- is tightly correlated with the energy of the particle. In spectroscopy applications, rf monochromators and energy selecting apertures are not mutually exclusive elements of the microscope column. Pairing our design with a downstream, large acceptance energy selector, which admits between 50\% and 100\% of beam current, offers the ability to tailor the tails of the final energy distribution and mitigates subleading sources of energy spread.

The body of this paper begins in Sec.~\ref{averagecurrent} with an analysis of the trade-off between current and energy spread in photoemission, and the statement of a fundamental lower bound on energy spread as a function of current on target. Considering photoemission from a planar source, Sec.~\ref{energyequalization} derives and solves analytic conditions on the cavity parameters for energy-spread minimization. Particle tracking simulations confirm these analytic results.  Precise synchronization is essential to minimising energy spread in our scheme and Sec.~\ref{jittersection} analyses the effect of timing jitter at the single femtosecond scale, precision that has been achieved with bunching cavities in ultrafast diffraction beam lines \cite{otto2017solving}. Section~\ref{brightness} investigates the effect of the cavities on the transverse coherence of the beam. Analytic results show that the same cavity parameters that are optimal for energy spread reduction also perfectly cancel the effects of spherical and chromatic aberrations in the electron gun. We compare this prediction of our analytical formula with particle tracking simulations.

%Particle tracking simulations complicate the picture, revealing a small trade-off between final energy spread and transverse coherence due to the non-impulsive deflection of particle trajectories during transit through the cavities. We foreshadow a method to correct for this higher-order loss of coherence with additional cavities, leaving the details to a future work. The analytic model and simulation tests of the design presented in this paper show that the loss of transverse coherence becomes negligible as source energies spreads approach $100 \ \mathrm{meV}$ and below.%

\section{Fundamental trade-off between energy resolution and beam current \label{averagecurrent}}

\begin{figure}
    \centering
    \includegraphics[width=1.1\linewidth]{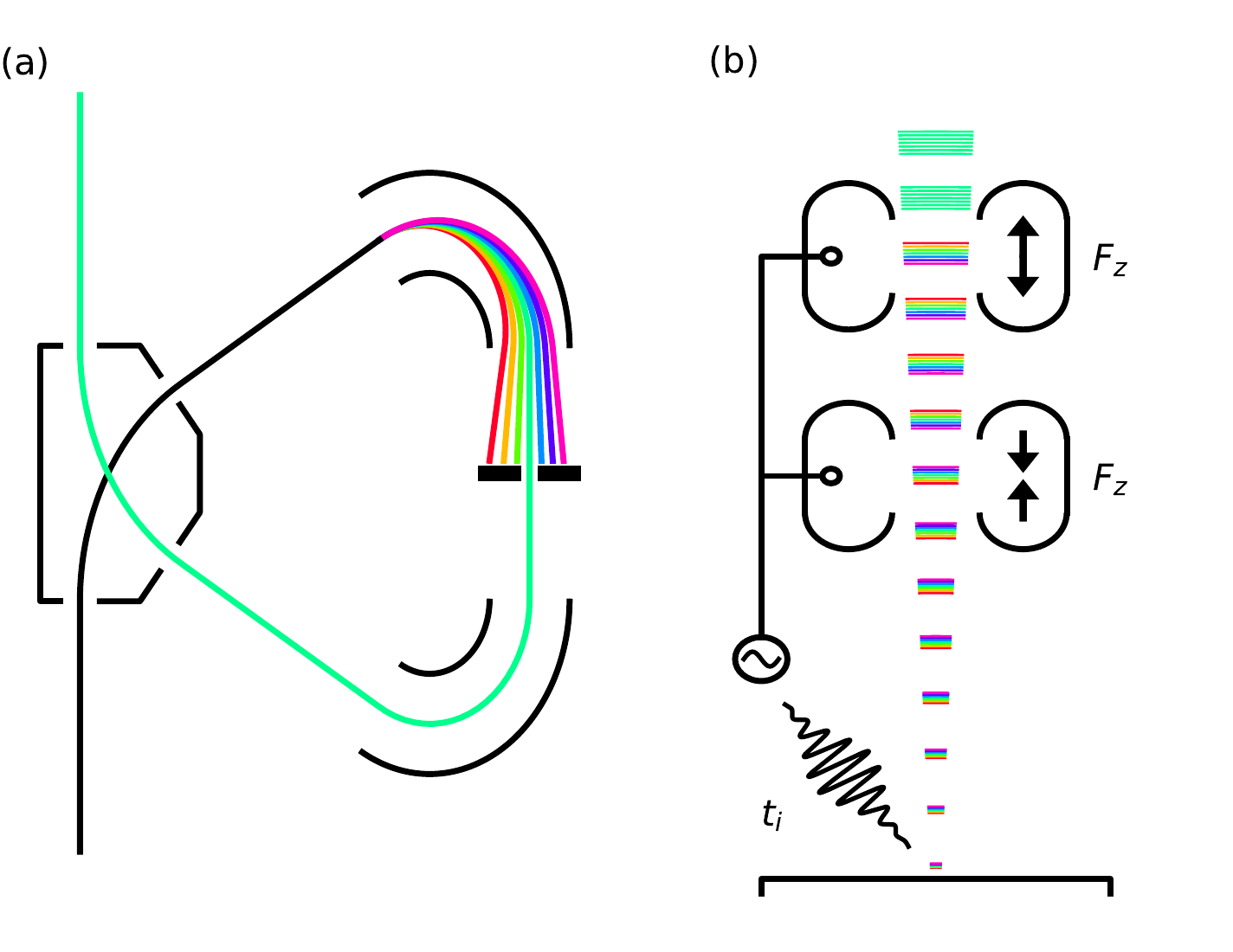}
    \caption{(a) Energy selecting monochromator as described in \cite{krivanek2009high}: magnetic prisms disperse the energy spectrum of the electron beam in the transverse direction and a narrow acceptance slit selects the desired bandwidth at the cost of lost current. (b) Our lossless monochromator design: photoemission is triggered by a laser pulse and the beam energy spectrum disperses in the longitudinal direction. Time-correlated acceleration in the pair of cavities, indicated by force vectors in the figure, equalizes the energies of the dispersed beam. The average current in our design is limited by the cavity frequency.}
    \label{sidebyside}
\end{figure}

 Monochromation entails a trade-off between final energy spread and average current on target, both in existing aperture-based energy-selectors and our proposed lossless design.  Figure~\ref{sidebyside} shows a schematic of our design side-by-side an energy-selector. The constraints that impose the current-energy trade-off are different between the two devices. A comparison helps to situate our design in relation to the state of the art. A first analysis is simplified by neglecting the contribution that transverse momenta make to total particle energy. The end of this section returns to the complication introduced by accounting for the transverse store of energy. 
 
 The conservation of longitudinal emittance in a pulsed beam relates the minimum energy spread achievable in lossless transport $\Delta {\cal E}_\mathrm{min}$ to the initial laser pulse length $\Delta t_l$ at the source, the final electron pulse length $\Delta t_f$, and the initial electron energy spread $\Delta K$:
\begin{equation}\label{lemittance}
    \Delta {\cal E}_{\text{min}} = \frac{\Delta t_l}{\Delta t_f}\Delta K \geq \frac{\hbar}{2 t_f}.
\end{equation}
The rightmost inequality is a consequence of the Heisenberg uncertainty principle, which sets the fundamental limit to longitudinal emittance. The factor $\Delta t_l$ includes the response time of the photocathode: on the scale of 10 fs \cite{Li2017} for typical metallic photocathodes, and extending much longer (up to 100 fs and above) for semiconductor photocathodes \cite{Karkare2014, Rao2014}. The single-electron-per-pulse regime reaches the lowest possible emittances (both transverse and longitudinal) because of the absence of Coulomb interactions that would otherwise broaden the energy distribution and spoil transverse coherence. At GHz or slower repetition rates, and primary energies of 10 keV or more, successive pulses are separated by distances greater than 1 cm, so that interactions between pulses are safely neglected. In this regime, an upper bound on the average current of a laser-driven system synchronized to rf cavities is $I_{av} = f e$, where $e$ is the electron charge and $f$ is the resonant frequency of the cavities.  Poisson emission statistics imply that a current less than this upper bound is required to adequately mitigate the effect of Coulomb interactions, but the difference is an order unity factor and appropriately neglected in the present scaling analysis. In time-resolved, pump-probe systems, the maximum practicable repetition rate is set by the time it takes the sample to relax to the ground state after pump excitation. Relaxation times vary significantly depending on the sample and the desired excitation strength.

\begin{figure}
    \centering
    \includegraphics[width=1.0\linewidth]{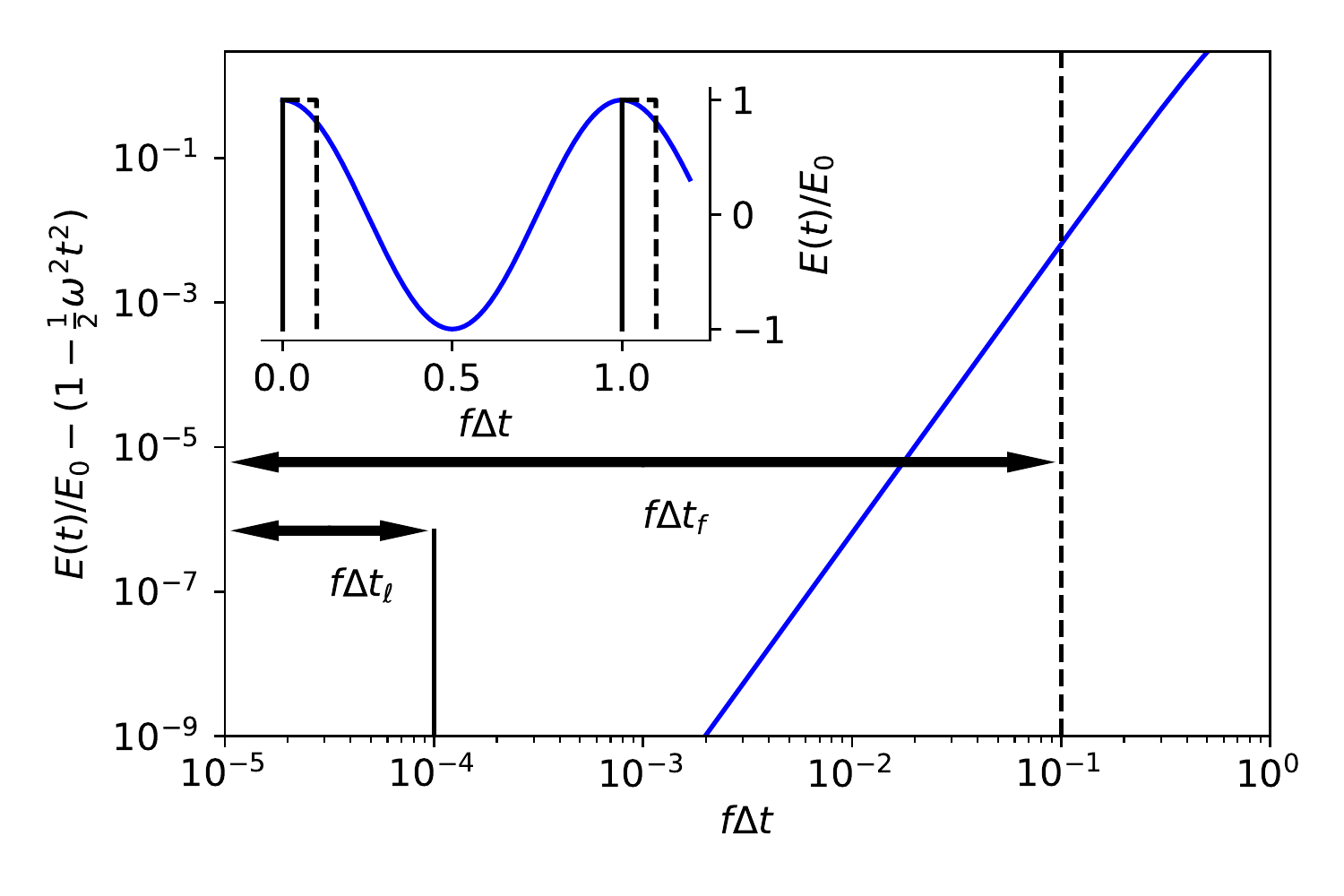}
    \caption{Normalized energy gain versus the single particle time of arrival, expressed as a fraction of the full rf period. The quadratic coefficient of the energy gain is subtracted, leaving only unwanted higher-order terms.  Vertical lines show the initial pulse length $\Delta t_\ell$ (solid) and final pulse length $\Delta t_f$ (dashed). The duty cycle in Eq.~\eqref{scaling} is $f\Delta t_f$.  Inset shows the full sinusoidal energy gain versus the scaled time of arrival, with the quadratic term restored.}
    \label{dutycycle}
\end{figure}

Pulses that stretch to fill the entire rf cycle acquire unwanted nonlinear energy-time correlations, and these higher-order effects thus bound the allowable final pulse length. Let the  \emph{duty cycle} $D$  denote  the ratio of final pulse length to rf period. Substituting the duty cycle into Eq.~\eqref{lemittance} yields an expression for our design's maximum average current:
\begin{equation}\label{scaling}
    I_{av} = \frac{D e}{\Delta t_l} \frac{\Delta {\cal E}_{\text{min}}}{\Delta K} = I_{pk}D \frac{\Delta {\cal E}_{\text{min}}}{\Delta K},
\end{equation}
where $I_{pk}:=e/\Delta t_l$ is the peak current at the cathode. Energy-selecting monochromators also show a linear scaling of average current with the fractional reduction in energy spread. The optimal performance of our monochromator is thus equivalent to an energy selector with an effective input current of $I_\mathrm{pk}D$. An estimate of the allowable values of $D$ depends on the details of our monochromator design. As shown in Sec.~\ref{energyequalization}, minimising final energy spread requires that the energy gain from the cavities depends quadratically on time. Figure~\ref{dutycycle} plots the cubic and higher-order time dependence of the work done by a single cavity on a logarithmic scale. Inspection of the figure indicates that choosing $D < 0.1 $ suppresses the higher-order contributions to the 1\% level or lower. Modern ultrafast laser oscillator sources can provide multiple GHz repetition rates with pulse durations well below $30$ fs \cite{Bartels1999}. For a total emission time of $\Delta t_l = 30$ fs, and assuming $D = 0.01$, we arrive at $I_{pk}D = 50$ nA, which compares favorably with the order $1$ nA current delivered from CFE sources to state-of-the-art TEM monochromators \cite{Krivanek2014}, of like design to the diagram shown in Fig.~\ref{sidebyside}(a). It bears repeating with respect to this example that 50 nA is not the average current input into the rf monochromator --- our rf design is lossless and so the average current is the same upstream and downstream, equal to $50 \ \mathrm{nA} \times \Delta {\cal E}_\mathrm{min} / \Delta K$.

Constrained by the transverse brightness of the electron source, energy selection in the spatial domain confronts a trade-off between energy and spatial resolution. The $10 \ \mathrm{pA}$ scale of the maximum current the CFE energy selector delivers to the experimental target is a consequence of balancing the objectives of angstrom scale spatial resolution against $10 \ \mathrm{meV}$ scale energy resolution \cite{krivanek2013monochromated}.  A CFE source delivering $1 \ \mathrm{nA}$ to the selector input, accelerated to $100 \ \mathrm{keV}$ at a brightness of $\mathrm{10}^{9} \ \mathrm{A} \ \mathrm{cm}^{-2} \ \mathrm{Sr}^{-1}$,  has a normalised transverse emittance of 1 pm, less than the Compton wavelength of the electron and a factor five greater than the fundamental lower bound set by the quantum uncertainty principle. There is thus room for increases in current on target with existing energy selector technology as the transverse emittance of the highest-resolution instruments approaches the quantum limit.

The rf monochromator presented in this paper balances an orthogonal trade-off between temporal resolution and energy resolution, constrained by longitudinal brightness.  Holding the pulse length and energy spread at the sample constant, a reduction in the initial pulse length results in higher current on the target. Compared to transverse brightness, there is far greater scope for future improvements in longitudinal brightness, given that the best ultrafast sources in use today --- at 10 to 100 fs pulse lengths and 100 meV energy spread --- are hundreds of times poorer than the fundamental brightness limit. To calculate the fundamental theoretical ceiling on the current that the rf monochromator can deliver to an experimental target, we assume an initial pulse length that approaches the quantum limit, $\Delta t_\ell \Delta K = \hbar /2$. Then, letting the duty cycle be $D=0.01$ and the final energy spread be $10 \ \mathrm{meV}$, the current on target is 20 nA, three orders of magnitude potential improvement. The implied repetition rate at an average current of 20 nA is 120 GHz, not practically feasible with present technology. Nevertheless, this example serves as a signpost of the ultimate, physical limits of rf monochromation in the one-electron-per-pulse regime.

Both metallic photoemission and cold field emission sources in use today produce energy spreads of multiple hundreds of meV \cite{Dowell2009, feist2017ultrafast, Krivanek2014, Scheinfein1993}. A final spread of 5 meV is an appropriate benchmark, being the resolution required to resolve phonons in EELS and to reduce chromatic aberration in objective lenses by more than an order of magnitude.  A target $\Delta {\cal E}_{\text{min}} $ of 5 meV from a $\Delta K = 500$ meV source with initial pulse length $\Delta t_l = 30$ fs implies a final pulse length of $3$ ps, a resonant cavity frequency of $\sim 3$ GHz,  and an average current of $500$ pA. An average current of $500$ pA is more than sufficient for imaging above 10 keV primary energy, and cavities and rf sources at $3$ GHz are well explored in both accelerator and time-resolved experimental work. Additionally, with the reduction in laser repetition rate (by pulse picking, for example) to accommodate sample recovery times in pump probe experiments, 3 ps resolution enables the tracking of phonon population evolution in time \cite{Stern2018}. 

In the applications of interest, the cathode makes the dominant contribution to energy spread, with the subleading contribution coming from fluctuations in the accelerating voltage. Our design specifically corrects source energy spread, relying on the correlation between {\em initial} kinetic energy and time and position of arrival. Section~\ref{jittersection}, on jitter, suggests how to incorporate into our design fast feedback from existing beam diagnostic devices so as to compensate subleading sources of energy spread.

The advantage of lossless monochromation over energy selectors for time-resolved applications is that users can obtain improved energy resolution without paying a cost in average current. For applications that at present obtain the best performance from CW beams, the scaling with peak current in Eq.~\eqref{scaling} points to the potential superiority of pulsed beams with lossless monochromation as higher-brightness photoemission sources become available. The two dimensions of active research toward higher-brightness photoemitters are lower source energy spread and smaller source size. Measurements of photoemission from cryo-cooled alkali-antimonide photocathodes have shown source energy spreads on the 10 meV scale \cite{cultrera2015cold,musumeci2018advances}, an order of magnitude smaller than CFE sources. Photoemitting tips yield nanometer source sizes, smaller than the diffraction limited laser spot diameter   \cite{feist2017ultrafast}. A hypothetical alternative to a tip geometry is to layer a photoemission mask on planar cathodes, exposing a photoemtting disc with a diameter on the scale of $10 \ \mathrm{nm}$. The simulation results we present in Sec.~\ref{energyequalization} make practical assumptions about the photoemission source that anticipate future trends. We consider a planar cathode geometry with an rms source size of $12 \ \mathrm{nm}$ and initial uniform energy spreads of $0.1, 0.5$ and $1$ eV. The physics that makes rf monochromation possible, which the next section describes, does not depend on assumptions about source quality.

The discussion in this section is completed by considering transverse spatial degrees of freedom. The contribution that the transverse momenta make to total energy spread implies a parallel trade-off between transverse beam size and energy spread. Therefore, including the transverse contribution makes possible a reduction in energy spread without a compensating increase in pulse length. A natural mechanism to imagine realizing this possibility is a radial electric field that performs work to reduce the energy stored in the transverse momentum and, as a byproduct, collimates the beam. Our design employs a similar mechanism, conceptually more complicated but simpler to engineer. Our design, after expanding the beam, applies a spatially varying longitudinal field so as to balance an energy surplus in the transverse direction by creating an energy deficit in the longitudinal direction. 

Altogether, our design achieves its final energy spread by stretching the beam both transversely and longitudinally. The spatial and temporal trade-offs do not originate from two distinct constraints, but are instead both consequences of the conservation of the six-dimensional phase space volume. Thus, if the expanded, reduced energy spread beam is again focused down to its source size or smaller by magnetostatic lenses, then the pulse length stretches as the beam size shrinks. Magnetostatic lenses do no work and hence have no effect on energy spread, but produce pulse stretching due to variations in the path lengths traced by particles.  In the applications of our monochromator previewed in the Introduction, there is more experimental flexibility to trade longer pulses for lower energy spread than there is to vary beam size at the experimental target. Hence, Eq.~\eqref{lemittance} formulated only in terms of the longitudinal variables captures the essential physics of our monochromator design.

\section{Energy equalization \label{energyequalization}}

\subsection{Qualitative summary}

Our monochromator design operates in two steps. First, the system of electron source and optical column correlates energy with time and position of arrival, stretching the electron pulse in time and space. The uncorrelated energy spread decreases as the pulse is stretched. In the second step, the rf fields remove the correlated energy spread.

\begin{figure}
    \centering
    \includegraphics[width=\linewidth]{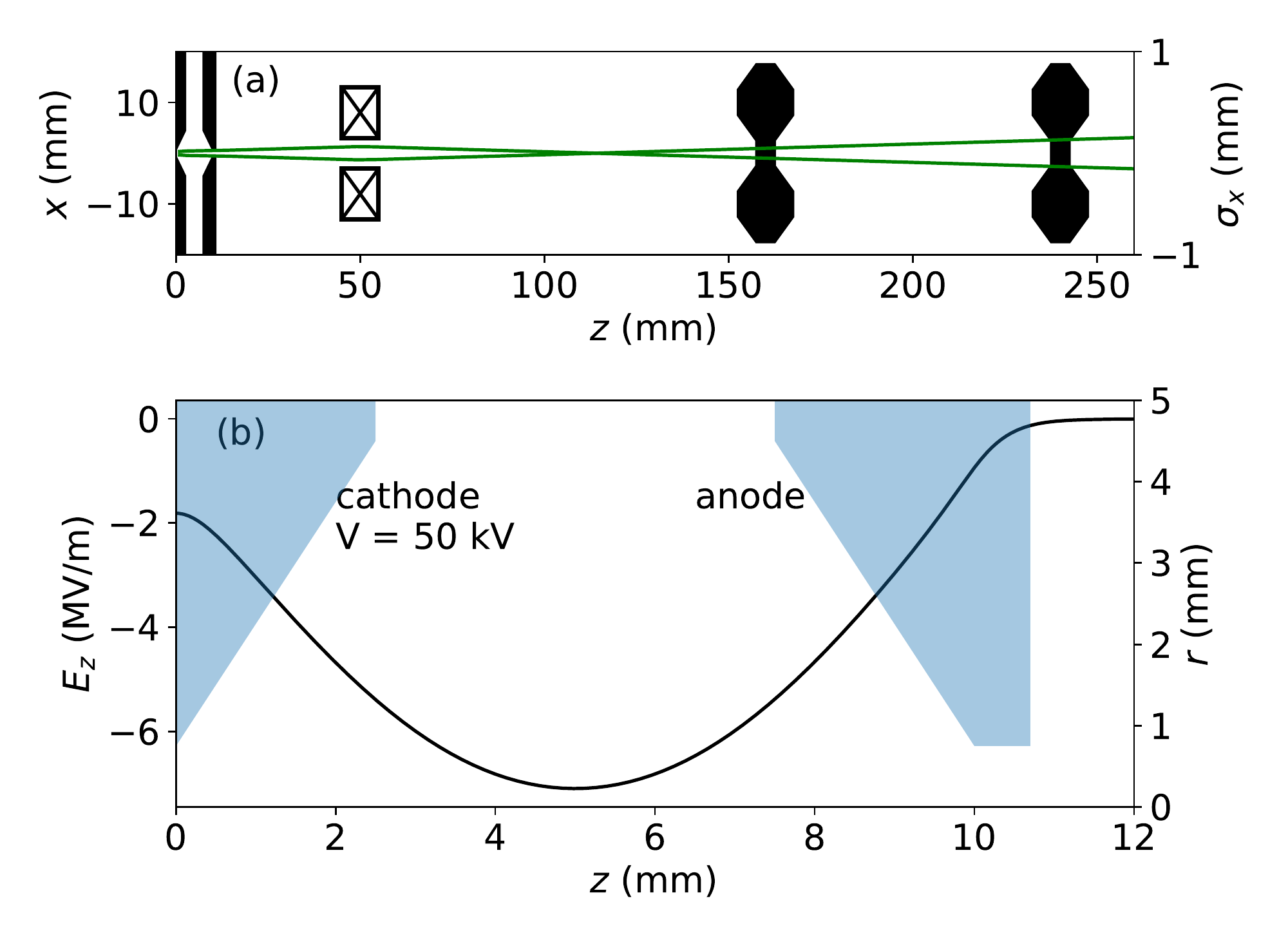}
    \caption{(a) Simulated source and monochromator layout in the $x$-$z$ plane. The left axis is the transverse scale for the optical elements.  All elements are axially symmetric. The transverse beam size is shown by the green curve, with the scale indicated by the axis on the right. The cathode and anode are at $z=0, \ 10 \ \mathrm{mm}$, respectively. At $z=50 \ \mathrm{mm}$  is a focusing solenoid; at $z=160, \ 240 \ \mathrm{mm}$, dumbbell silhouettes approximate cavity cross sections. (b) Electrode profile and axial field $E_z(z)$ in the gun for a voltage of 50 kV.}
    \label{gun}
\end{figure}

\begin{figure*}
    \centering
    \includegraphics[width=\linewidth]{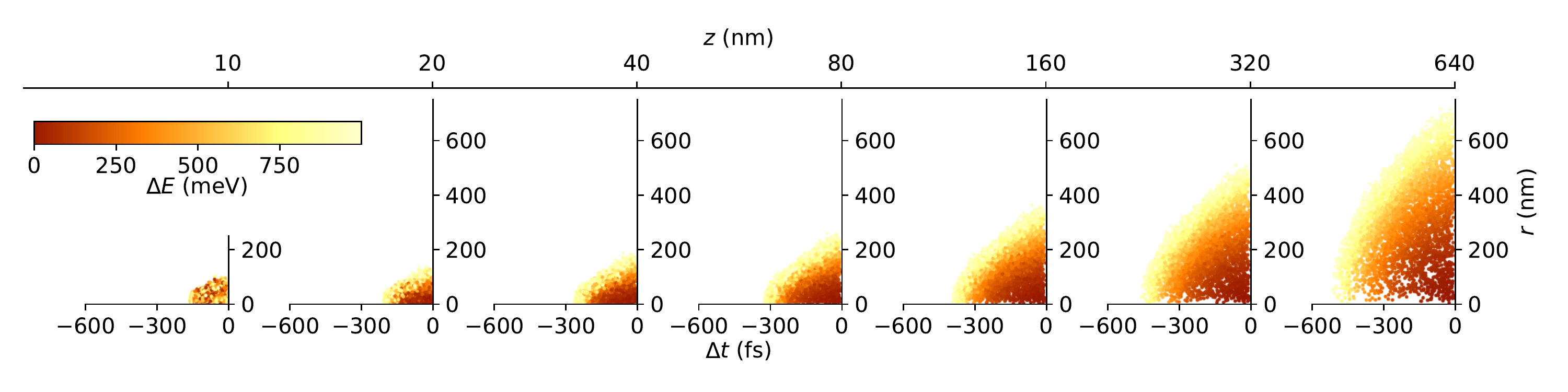}
    \caption{Evolving correlation between the energy, radial position $r$, and time of arrival $\Delta t$ in a dc electron gun. Results plotted are of single particle trajectory simulations. The accelerating field is uniform with a gradient of $5 \ \mathrm{MV/m}$. Each subplot shows the cross section of radial position and time of arrival at the exponentially increasing values of $z$ indicated on the top axis. Color indicates particle energy $\Delta {\cal E}$ relative to the minimum energy in the statistical ensemble; $\Delta t$ is defined relative to the last arriving particle. As the uncertainty in time and position grows, the correlation with energy tightens. Uncertainty in time of arrival asymptotically approaches the value predicted by Eq.~\eqref{pz0} as the energy gained in acceleration comes to dominate the initial kinetic energy. The initial statistical ensemble in all our simulations, unless otherwise specified, is uniformly distributed in energy, time of emission, and solid angle over the forward hemisphere \cite{Dowell2009}.}
    \label{filter}
\end{figure*}

Correlations between energy, time and space naturally arise near zero energy in the low-to-moderate extraction fields ($ 1 - 10$ MV/m) of a dc (i.e. electrostatic) gun, which our design exploits. Figure~\ref{gun} shows the field-map used to simulate the electron gun alongside a plot of the position of downstream beamline elements. Illustrative snap shots of the evolution of energy correlations in simulation are shown in Fig.~\ref{filter}. An alternative method for stretching electron pulses is to insert a dedicated device comprising rf cavities and drift space \cite{Franssen2017, zeitler2015linearization}. Our omission of dedicated stretching cavities has the advantage of simplicity. Furthermore, by relying on the gun field to stretch the pulse, our design does not require significant drift space for stretching to develop. In this section, we derive an analytical model of stretching in an idealized gun field and confirm the model's accuracy in particle tracking simulations. 

\begin{figure}
\includegraphics[width=0.5\textwidth]{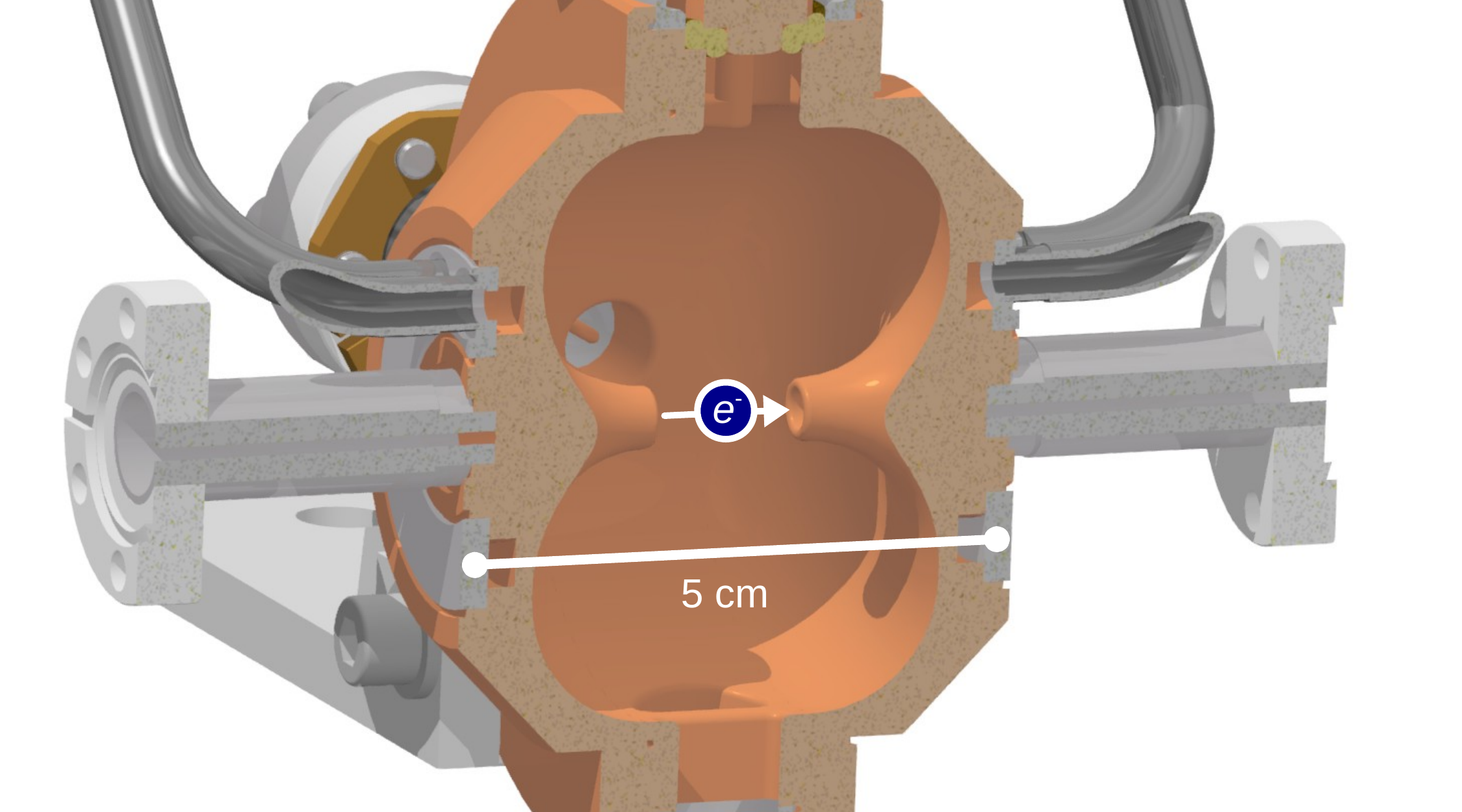}
\caption{\label{buncher} Cross section of the $3 \ \mathrm{GHz}$ TM010 mode cavity used in the energy equalization device described in this paper.}
\end{figure}

Two rf cavities correct the kinetic energy spread contained in the time (cavity one) and transverse position (cavity two) degrees of freedom. Both cavities are identical in design, comparable to the device described in \cite{pasmans2013microwave}. A three-dimensional rendering of our cavity is shown in Fig.~\ref{buncher}.  This section computes to good accuracy the settings of cavity phases and amplitudes that minimize the final energy spread. 

Simply knowing the initial kinetic energy of a particle does not uniquely determine its time and position of arrival at the cavities. Instead, we derive the inverse relationship: the initial kinetic energy as a function of the time and position of arrival at the cavities. The exit of the gun serves as the primary reference plane for particle coordinates. Any equipotential plane downstream of the anode would fulfil this role equally well. Our analytic model ignores fringing effects and thus takes the anode to be the gun exit. The initial kinetic energy is nonrelativistic and thus proportional to the sum of squares of initial momenta. Our problem thus reduces to expressing each of the three initial momenta separately in terms of their conjugate coordinates at the gun exit. 

\subsection{Correlation between emission energy, arrival time, and position}

We begin by deriving the initial longitudinal momentum of a particle as a function of time of arrival at the reference plane, the gun exit. For our analytic derivation, we assume that the gun is a uniform field of strength $E^\mathrm{cat}_z \hat{\bf z}$. We relax this assumption later. As is standard in electron optics, we take longitudinal position $z$ to be the independent variable of electron motion and treat time of arrival $t$ as a function of $z$. Differences in times of arrival $\Delta t$, as measured by a clock at rest in the laboratory frame, play a crucial role in the derivation and should not be mistaken for time as measured in a frame comoving with the beam. Our reference frame is at rest with respect to the laboratory throughout. The relativistic equations of motion for a charged particle in a uniform electric field $E^\mathrm{cat}_z \hat{\bf z}$, solved for $t(z)$, give,
\begin{equation}\label{time}
t(z) = t_0 + \frac{p_{z0}}{eE_z^\mathrm{cat}} +  \sqrt{\frac{z^2}{c^2} - \frac{2z\gamma_0m_e}{eE_z^\mathrm{cat}} + \frac{p_{z0}^2}{(eE^\mathrm{cat}_z)^2}},
\end{equation}
where $t_0$ is the time of emission, $p_{z0}$ is the initial longitudinal momentum at the cathode, and $\gamma_0$ is the initial relativistic energy factor at the cathode. Three terms add in quadrature under the square root, and we name them for convenience. First, the {\em ultra-relativistic time of flight}, $z/c$, which is the time it would take a photon to travel the distance to $z$. Next, the {\em Newtonian time of flight} $\sqrt{-2zm/eE^\mathrm{cat}_z}$ is the time it would take a nonrelativistic particle of mass $m=\gamma_0m_e$ to reach $z$, starting at rest. Finally, the {\em stopping time} $-p_{z0}/{eE^\mathrm{cat}_z}$ is the time it would take the accelerating gradient to stop a particle fired toward the cathode with initial momentum ${\bf p} = -p_{z0}\hat{\bf z}$.

In the regime of interest --- particles with $\sim 1$ eV initial energies and $>1$ keV final energies --- the smallest term is the stopping time. If we have $E^\mathrm{cat}_z = -1 \ \mathrm{MV/m}$, the stopping time of an electron with velocity $ v_z/c = 0.002 $ is $3 \ \mathrm{ps}$, the ultra-relativistic time of flight across a cathode gap of $1 \ \mathrm{cm}$ is $30 \ \mathrm{ps}$, and the corresponding Newtonian time of flight is $500 \ \mathrm{ps}$.

In order to estimate the error that arises from neglecting the stopping time, we can expand the square root appearing in Eq.~\eqref{time} in powers of the small quantity
\begin{equation}\label{delta}
    \delta = \frac{p_{z0}^2}{\left(eE^\mathrm{cat}\right)^2}\left(\frac{z^2}{c^2} - \frac{2z\gamma_0m_e}{eE^\mathrm{cat}_z}\right)^{-1},
\end{equation}
 which is proportional to the square of the stopping time. Then, for the example values cited above, the correction term first order in $\delta$ is $\sim 10 \ \mathrm{as}$. The relevant scale to compare is the uncertainty in the time of emission of a single electron packet $\Delta t_l$, equal to the the laser pulse length and photoemission response time, which are on the order of $10 \ \mathrm{fs}$. Thus, the stopping time can safely be neglected inside the square root. Under the assumption that all initial kinetic energies are of single eV scale or less, it also safe to assume that $\gamma_0 = 1$, and no remaining term inside the square root appearing in Eq.~\ref{time} depends on the initial conditions. Thus, the square root drops out from the difference in arrival times $\Delta t$ between two particles, and $\Delta t$ becomes linearly proportional to the difference in their initial longitudinal momenta $\Delta p_{z0}$,
\begin{equation}\label{pz0}
    \Delta t = \frac{\Delta p_{z0}}{eE^\mathrm{cat}_z} + \Delta t_0,
\end{equation}
with $\Delta t_0$ the difference in emission times (relative to the arrival time of the laser pulse).

The most convenient choice of reference particle is the particle with zero initial kinetic energy. Arrival time differences $\Delta t$ therefore give us the $p_{z0}$ of all particles up to the precision set by $\Delta t_l$, which is $1 \%$ for the example values. 

The coefficient of proportionality appearing in Eq.~\eqref{pz0} is independent of $z$ and thus the gun length. A dependence on gun length is absent because the relative time of arrival is frozen for $z$ large enough that $\delta \ll 1$, per Eq.~\eqref{delta}. To describe this freezing effect more explicitly, consider two particles with longitudinal velocities $\beta_{z1}(z)c, \ \beta_{z2}(z)c$ that differ in arrival time by $\Delta t (z_1)$ at longitudinal position $z_1$. The difference in arrival time of the same two particles at a downstream location $z_2$ is
\begin{equation}\label{explicit}
   \Delta t (z_2) = \Delta t(z_1) + \int_{z_1}^{z_2}\frac{\left[\beta_{z2}(z)-\beta_{z1}(z)\right]}{\beta_{z1}(z)\beta_{z2}(z)}\frac{dz}{c}.
\end{equation}
Now suppose an accelerating gradient such that as $z_1$ goes further downstream, the greater is the mean particle energy. It then follows that, for $z_1$ sufficiently far from the cathode, the integrand on the right-hand side of Eq.~\eqref{explicit} vanishes and $\Delta t$ becomes independent of $z$, no matter the functional dependence of velocity on position. The freezing of time of arrival differences makes it possible to generalize Eq~\eqref{pz0} to nonuniform accelerating fields by replacing $E^\mathrm{cat}_z$ with the photocathode field. The resulting expression is accurate so long as (i) the photocathode field is approximately uniform over a distance $z$ such that $\delta \ll 1$ and (ii) the particle velocities are increasing functions of $z$ for the remaining length of the gun.

Fluctuations in the accelerating gradient,  typically at the $10^{-5}$ level or below, affect electron time of flight and are a subleading source of uncertainty in the correlation between $\Delta t$ and $p_{z0}$. To estimate the significance of this effect, we expand Eq.~\eqref{time} to linear order in the gradient fluctuation $\Delta E^\mathrm{cat}_z$. Letting $z_\mathrm{gun}$ be the location of the gun exit and $z$ a location in the drift region downstream of the gun, the resulting change in arrival time $\Delta t_\mathrm{gun}(z)$ is,
\begin{align}
    \Delta t_\mathrm{gun}(z) =& -\frac{1}{4}\frac{\Delta E_z^\mathrm{cat}}{E_z^\mathrm{cat}}\left(1+\frac{z}{z_\mathrm{gun}}\right) \sqrt{- \frac{2z_\mathrm{gun}m_e}{eE_z^\mathrm{cat}}}.\label{gunjitter}
\end{align}
The expression in Eq.~\eqref{gunjitter} neglects relativistic terms, which, if included, further suppress the size of $\Delta t_\mathrm{gun}$. The estimate for $E_z^\mathrm{cat} = -1 \ \mathrm{MV}/\mathrm{m}$ and $z_\mathrm{gun} = 1 \ \mathrm{cm}, \ z = 10 \ \mathrm{cm}$ is $\Delta t_\mathrm{gun}(z) < 10 \ \mathrm{fs}$, less than the dominant contribution to time of flight uncertainty from the emission process $\Delta t_l$.

As for the transverse coordinates, the solution to the equation of motion in $x$ (without transverse focusing) is,
\begin{equation}
    x = \frac{p_{x0}}{m_e}\tau +x_0,\label{eomX}
\end{equation}
with $x_0$ the coordinate of the particle at emission and $\tau$ the proper time that elapses between emission of the particle and its crossing the transverse plane at $z$. The coordinate $y$ follows from cylindrical symmetry. Let $\tau_*(z)$ be the proper time to reach $z$ of the reference particle (initially at rest). Expanding $\tau$ in $p_{z0}$ around $\tau_*(z)$,
\begin{align}
    \tau &= \tau_* + \frac{d\tau}{dp_{z0}}p_{z0} + {\cal O}\left(p_{z0}^2\right) \notag \\
    &= -\frac{m_ec}{eE^\mathrm{cat}_z}{\cal A}({\gamma_*}) + \frac{p_z}{eE^\mathrm{cat}_z} + {\cal O}\left(p_{z0}^2\right). \label{tau}
\end{align}
Here, the variable $\gamma_*(z)$ is the relativistic factor for the reference particle and,
\begin{equation}\label{aspect}
    {\cal A}(\gamma_*):=\ln\gamma_* +\ln\left(1 + \beta_*\right),
\end{equation}
with $\beta_*(z)c$ the velocity of the reference particle. Dividing Eq.~\eqref{eomX} by Eq.~\eqref{pz0} shows that ${\cal A}$ approximates the aspect ratio $x/c\Delta t$; hence, the choice of notation.

To find the initial kinetic energy $K$ as a function of a particle's position and time of arrival, we simply square previous expressions derived for momenta, i.e.,
\begin{equation}\label{K0}
    K(\Delta t,x,y) = \frac{(eE^\mathrm{cat}_z)^2}{2m_e}\left[(\Delta t - t_0)^2 + \frac{(x-x_0)^2 + (y-y_0)^2}{c^2{\cal  A}^2}\right],
\end{equation}
recalling that Eq.~\eqref{aspect} defines the aspect ratio ${\cal A}$ and that $E_z^\mathrm{cat}$ is the cathode field.
To arrive at this result, we make the approximation $\tau=\tau_*$. Simulation results verifying Eq.~\eqref{K0} are presented below in Sec.~\ref{energyequalization} D. The leading correction to Eq.~\eqref{K0} comes from considering the linear order in the expansion of $\tau$ shown in Eq.~\eqref{tau}, which contributes a cubic term to the right-hand side of Eq.~\eqref{K0},
\begin{equation}\label{cubiccorrection}
   \frac{\partial^3 K}{\partial t \partial r^2}  = \frac{2(eE^\mathrm{cat}_z)^3}{m_e^2c^3{\cal A}^3}.
\end{equation}
The relative size of this cubic term is,
\begin{equation}\label{a3}
           \frac{\partial^3 K}{\partial t \partial r^2} \Delta t \bigg/
           \frac{\partial^2 K}{\partial r^2} 
           = \frac{2\gamma_0\beta_{z0}}{\ln \gamma_* +\ln\left(1 + \beta_*\right)} \sim \frac{\beta_{z0}}{\beta_*}.
\end{equation}
The initial longitudinal particle velocity $\beta_{z0}$ is on the order $10^{-3}$ for electrons emitted with kinetic energies less than $1 \ \mathrm{eV}$. The final velocity $\beta_* \approx 0.5$ at an accelerating voltage of $50 \ \mathrm{kV}$. Thus, for pulse stretching factors approaching 1000 or more, cubic-order correlations put a floor on the final energy spread achievable with our two-cavity solution at parts per thousand the initial energy spread.

\begin{figure}
    \centering
    \includegraphics[width=\linewidth]{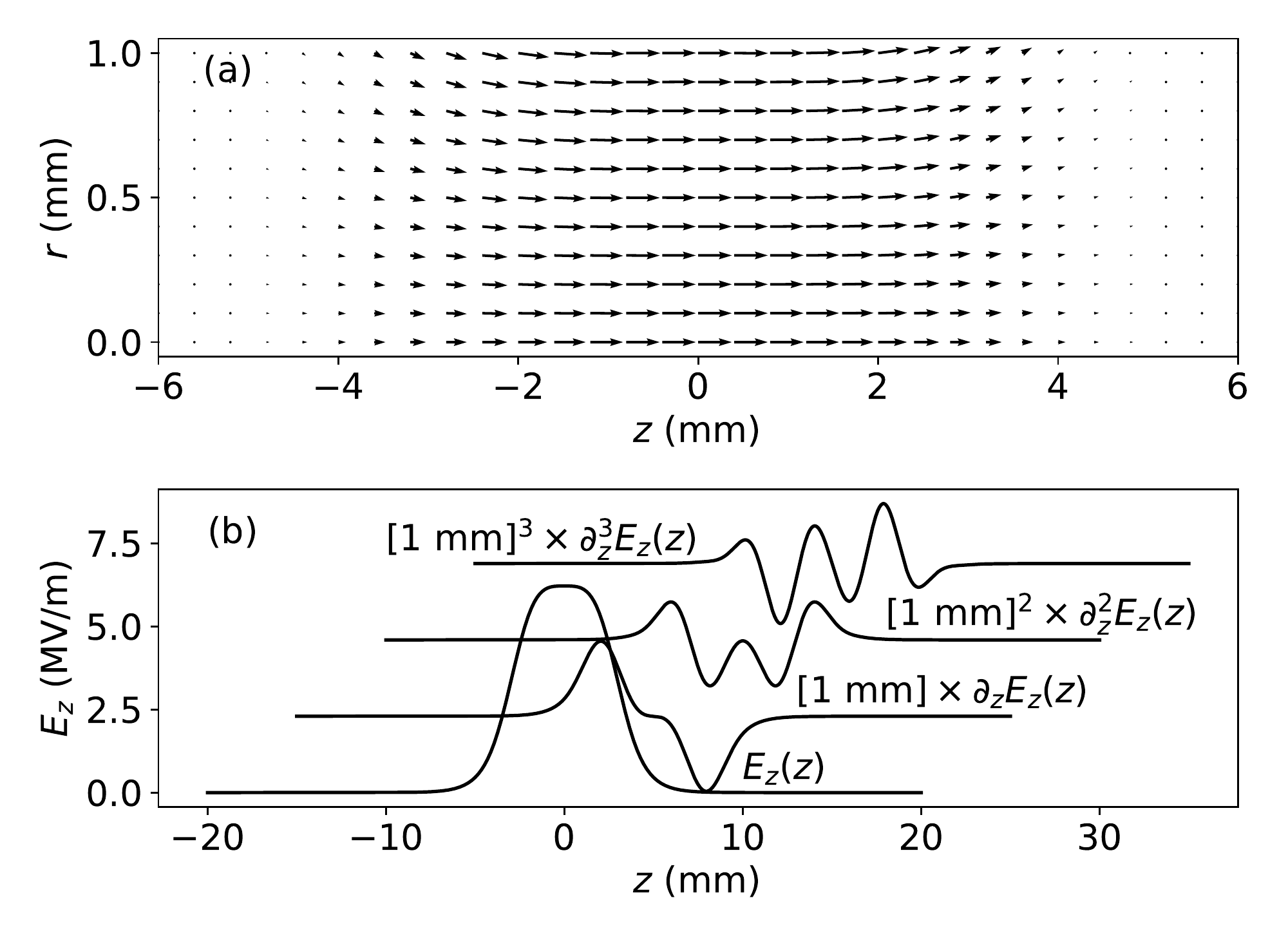}
    \caption{Simulation cavity field map: (a) $z, r$ cross section of the radially symmetric cavity ${\bf E}^\mathrm{rf}$ field at nominal amplitude and phase; (b) the axial cavity field $E^\mathrm{rf}_{0,z}$ and its leading spatial derivatives, from which we construct the simulation field map.}
    \label{allfields}
\end{figure}

\subsection{Cavity fields}
Knowing the initial kinetic energy of a particle as a function of the coordinates $\Delta t, x, y$ in the transverse plane at $z$, we must derive the matching expression for the work done by the cavity fields on transiting particles. For the purpose of this derivation, we make the following rigid-beam approximation, which is exact in the limit that the change in momentum due to the cavity field is small compared to the mean particle momentum. Namely, we assume that particles transit with constant velocity $\beta c$ parallel to the symmetry axis. In this limit, only the longitudinal component of the cavity electric field contributes to the work done. Figure~\ref{allfields} shows the electric field of the TM010 mode supported by our cavity design.

The longitudinal component of {\em any} axially symmetric transverse magnetic mode expands in powers of derivatives of the on-axis longitudinal field $E^\mathrm{rf}_{0,z}$ as,
\begin{equation}\label{fields}
     E^\mathrm{rf}_z(t,z,r) = J_0\left[ r \sqrt{\frac{\partial^2}{\partial z^2} + \frac{\omega^2}{c^2}}\right]E^\mathrm{rf}_{0,z}(z)\cos\left(\omega t + \phi_0\right),
\end{equation}
where $J_0$ is the zeroth-order Bessel function, which for operator arguments is defined by the power series,
\begin{equation}
    J_0\left[\frac{\partial}{\partial z}\right] := \sum_{k=0}^\infty
\left(\frac{-1}{4}\right)^k\frac{1}{(k!)^2}\frac{\partial^{2k}}{\partial z^{2k}}.
\end{equation}
 We define the free parameter $\phi_0$ such that at $\phi_0 = 0$ the reference particle undergoes the maximum change in energy. Appendix~\ref{cavexp} presents a derivation of Eq.~\eqref{fields}.
Integrating the right-hand side of Eq.~\eqref{fields} by parts to all orders in the derivative expansion gives the work $W$ as a function of the radial coordinate $r=\sqrt{x^2 + y^2}$ and the time of arrival $t$,
\begin{align}
W(r,t) = -\int_{-\infty}^{\infty} e{\bf E}^\mathrm{rf} \cdot d {\bf s} \approx -\int_{-\infty}^{\infty} eE^\mathrm{rf}_z(t(z), z, r) dz \notag \\
= -J_0\left( \frac{2\pi i}{\beta\gamma}\frac{r}{\lambda}  \right)\cos(\phi)\int_{-\infty}^\infty eE^\mathrm{rf}_{0,z}(z) \cos\left(\frac{2\pi}{\beta}\frac{z}{\lambda}\right)dz, \label{work}
\end{align}
with $\phi = \omega\Delta t + \phi_0$ and  $\Delta t$ the difference in time of arrival at the gun exit, the same variable appearing in Eq.~\eqref{K0}. Referring the time of arrival to the gun exit in this way is an approximation valid in the limit that drift sections between elements after the anode are short, which is the regime of interest. The intermediate steps in performing the integrations by parts are presented in Appendix~\ref{intparts}. The integrand on the right-hand side of Eq.~\eqref{work} is independent of $r$ and $t$. Hence, neglecting the small spread in $\beta$, it is possible to define an effective cavity length $d_{rf}$ and voltage \cite{pasmans2013microwave}. Letting $E^\mathrm{rf}_\mathrm{max}$ be the maximum accelerating field on axis,
\begin{equation}
    d_\mathrm{rf}: = \frac{1}{E^\mathrm{rf}_\mathrm{max}}\int_{-\infty}^\infty E^\mathrm{rf}_{0,z}(z) \cos\left(\frac{2\pi}{\beta_*}\frac{z}{\lambda}\right)dz.
\end{equation} %should be E_\mathrm{min}?%
 Choosing $\phi_0=0$ simplifies the conditions for canceling the quadratic dependence of energy on time appearing in Eq.~\eqref{K0}. With $\phi_0=0$, the expression for $W$ expands to quadratic order in $r$ and $t$ as,
\begin{equation}\label{cavity}
    W(r,t) = -d_\mathrm{rf}eE^\mathrm{rf}_\mathrm{max}\left(1 - \frac{1}{2}\omega^2 \Delta t^2 + \frac{\pi^2}{\beta^2\gamma^2}\frac{r^2}{\lambda^2} \right).
\end{equation}
The term in $r$ enters Eq.~\eqref{cavity} with the opposite sign to its counterpart in $t$ because Eq.~\eqref{fields} requires the peak of $E^\mathrm{rf}_{z,0}(z)$  to be a saddle point in three dimensions.

\subsection{Analytic prediction of cavity parameters}
By incorporating two cavities in the monochromator, it is possible to make the cumulative energy gain a concave down quadratic function of both time and space. A negative quadratic dependence cancels the positive dependence imprinted at the electron source. This strategy introduces the design problem of optimising beam transport between the cavities. It is conceptually simplest to suppose transfer optics that image the reference plane of the gun exit successively at the mid planes of the two cavities. If the imaging condition is satisfied, the transfer map that relates the particle transverse coordinates at the gun exit to the particle coordinate at a given cavity midplane is described by a single parameter, namely, the magnification factor $M, \ x \mapsto Mx $. Let $M_1$ and $M_2$ denote the magnification factors at each respective cavity, and let $E^\mathrm{rf}_{\mathrm{max},1}$ and $E^\mathrm{rf}_{\mathrm{max},2}$ be the two cavity amplitudes. It follows from Eqs.~\eqref{K0} and \eqref{cavity} that the difference in energy $\Delta {\cal E}$ between any particle and the reference particle, referred to the particle coordinates at the plane of the gun exit, splits into the two terms
\begin{equation}\label{allE}
    \Delta {\cal E} (t, r) = \Delta {\cal E}(t) + \Delta {\cal E}(r),
\end{equation}
where, ignoring terms in $t_0, x_0$ and $y_0$, as well as longitudinal drift,
\begin{align}
    \Delta {\cal E}(t) =& \left[\frac{(eE^\mathrm{cat}_z)^2}{2m_e} +d_\mathrm{rf}e\left(E^\mathrm{rf}_{\mathrm{max},1}- E_{\mathrm{max},2}^\mathrm{rf}\right)\frac{\omega^2}{2}\right]\Delta t^2, \label{tcav}\\
    \Delta {\cal E}(r) =& \left[\frac{(eE^\mathrm{cat}_z)^2}{2m_ec^2{\cal A}^2} \right. \notag \\
    &\left.-\frac{\pi^2d_\mathrm{rf}e}{\beta_*^2\gamma_*^2\lambda^2}\left(E^\mathrm{rf}_{\mathrm{max},1}M_1^2- E^\mathrm{rf}_{\mathrm{max},2}M_2^2\right)\right]r^2 \label{rcav}.
\end{align}
Again, $\beta_*$ and $\gamma_*$ are the values for a particle initially at rest. To optimize the cavities for monochromation, we set $\Delta {\cal E}(t)$ and $\Delta {\cal E}(r)$ to zero, yielding an under-constrained system of equations in the free variables $E^\mathrm{rf}_{\mathrm{max},1,2}$ and $M_{1,2}$. That the system is under constrained suggests the freedom to optimize a second design objective beyond energy spread. The most relevant second objective for microscopy applications is transverse emittance preservation, which we consider below in Sec.~\ref{brightness}. Though under-constrained, Eq.~\eqref{tcav} does predict the net energy added to the beam by the monochromator. The net energy added is a self-consistency test on the assumption that particle velocities remain approximately constant as they transit the cavities, which was made in deriving Eq.~\eqref{rcav}. The total energy gain of the beam is, 
\begin{equation}\label{diff}
-ed_{\mathrm{rf}}\left(E^\mathrm{rf}_{\mathrm{max},1}- E^\mathrm{rf}_{\mathrm{max},2}\right) = \frac{1}{4\pi^2}\frac{\left(eE^\mathrm{cat}_z\lambda\right)^2}{m_ec^2}.
\end{equation}
Given a $3 \ \mathrm{GHz}$ cavity and an accelerating gradient of $1 \ \mathrm{MV/m}$, the total energy added is $500 \mathrm{eV}$. This energy scale indicates that rf amplitude noise at the $10^{-5}$ level or larger makes a non-negligible contribution to the final energy spread.

Equation ~\eqref{tcav} ignores the uncertainty in emission time $t_0$ set by the laser pulse length and the photocathode response time $\Delta t_l$. A nonzero laser pulse length imposes another limit on the lowest achievable energy spread. If we take the laser pulse length into account then there is a linear correlation between emission time and energy in the final longitudinal distribution. The associated energy spread induced by nonzero emission time is,
\begin{equation}\label{residual}
    \Delta {\cal E}(t_0) = d_\mathrm{rf}e\left(E^\mathrm{rf}_{\mathrm{max},1} - E^\mathrm{rf}_{\mathrm{max},2} \right)\omega^2\frac{p_{z0}}{eE^\mathrm{cat}_z} t_0.
\end{equation}
Substituting the expression for the net energy gain at the optimal cavity parameters into Eq.~\eqref{diff} yields
\begin{equation}\label{residual2}
    \Delta {\cal E}(t_0) = -\frac{eE^\mathrm{cat}_z}{m_e}p_{z0}t_0.
\end{equation}
 Equation \eqref{residual2} shows that the final energy spread is bounded by the longitudinal emittance at the source, i.e., by the product of the spread in initial longitudinal momentum and time of emission. For the residual energy spread associated with non-zero initial transverse size $\Delta {\cal E}(r_0)$ to be of the same magnitude as or smaller than the residual $\Delta {\cal E}(t_0)$, the source transverse size must be on the scale of single microns, provided beam sizes in the cavities of $\sim 100$ microns. Source sizes for the highest brightness microscopy applications, both pulsed and continuous wave, are on the nanometer scale, and the simulation results presented in Sec.~\ref{energyequalization}~E assume a nanometer-scale source. Thus, we can neglect $\Delta {\cal E}(r_0)$ in estimating the final energy spread. A subtle implication of Eq.~\eqref{residual2} is then that, even for initial momenta with a statistical distribution in solid angle that spans the whole forward hemisphere, only the marginal distribution over $p_{z0}$ determines the final energy spread. Marginalising over solid angles tends to weight the marginal $p_{z0}$ distribution toward $p_{z0}=0$, reducing the final energy spread.

\subsection{Simulation results}

\begin{figure}
    \centering
    \includegraphics[width=\linewidth]{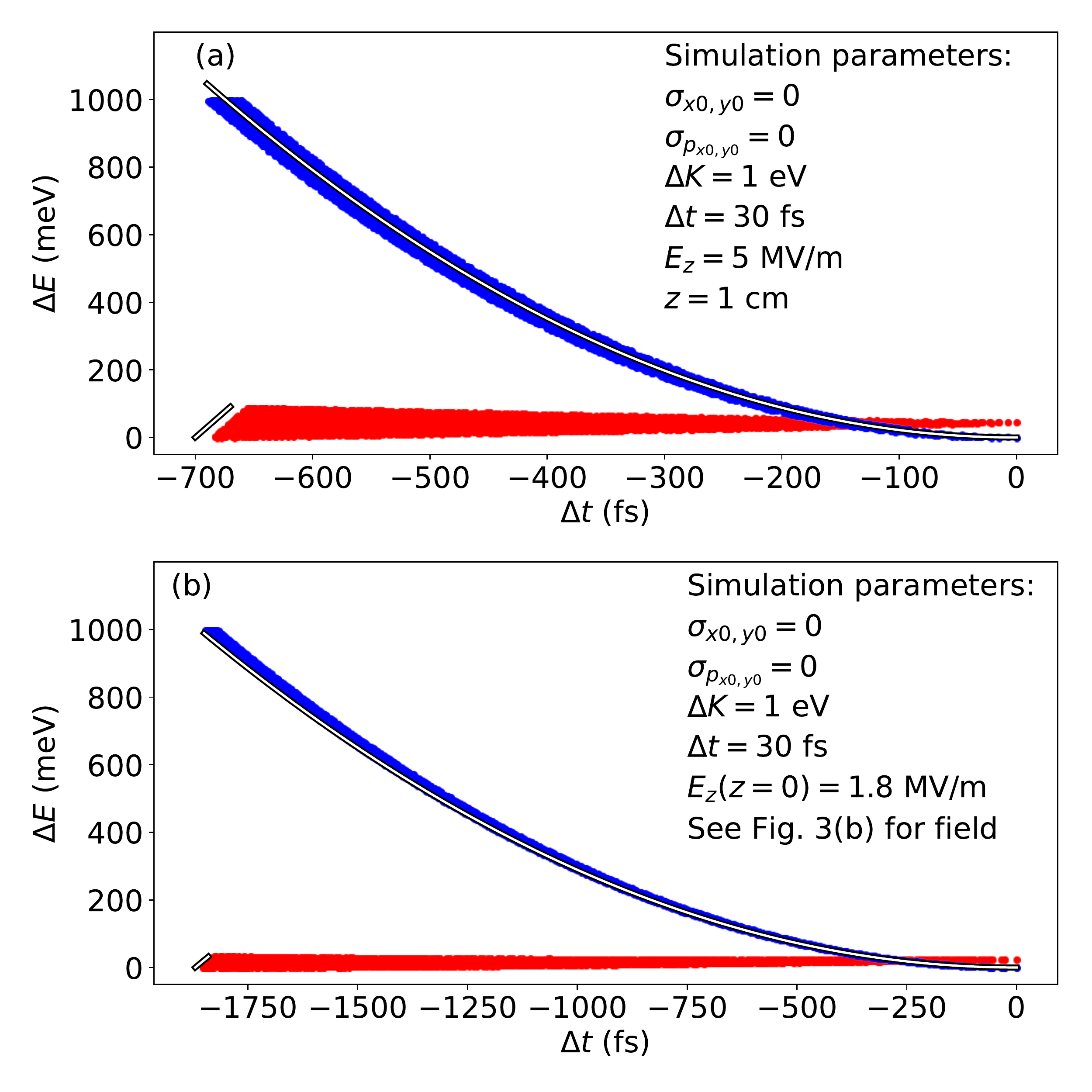}
    \caption{ Simulation of a beam with $\Delta t_l = 30$ fs and only longitudinal momentum spread. Transverse size and momentum spread are set to zero. In (a), an idealized, uniform field gun is used. In  (b), the realistic gun field shown in Fig.~\ref{gun} is used. Blue points denote the energy-time correlation just after the gun. Red points show the result after transiting a single cavity with amplitude set by the solution of solution of Eq.~\eqref{tcav}. White lines are the predictions of Eq.~\eqref{K0} and Eq.~\eqref{residual2}, where in (b) we use the field at the cathode for $E^\mathrm{cat}_z$.}
    \label{oneDt}
\end{figure}

\begin{figure}
    \centering
    \includegraphics[width=\linewidth]{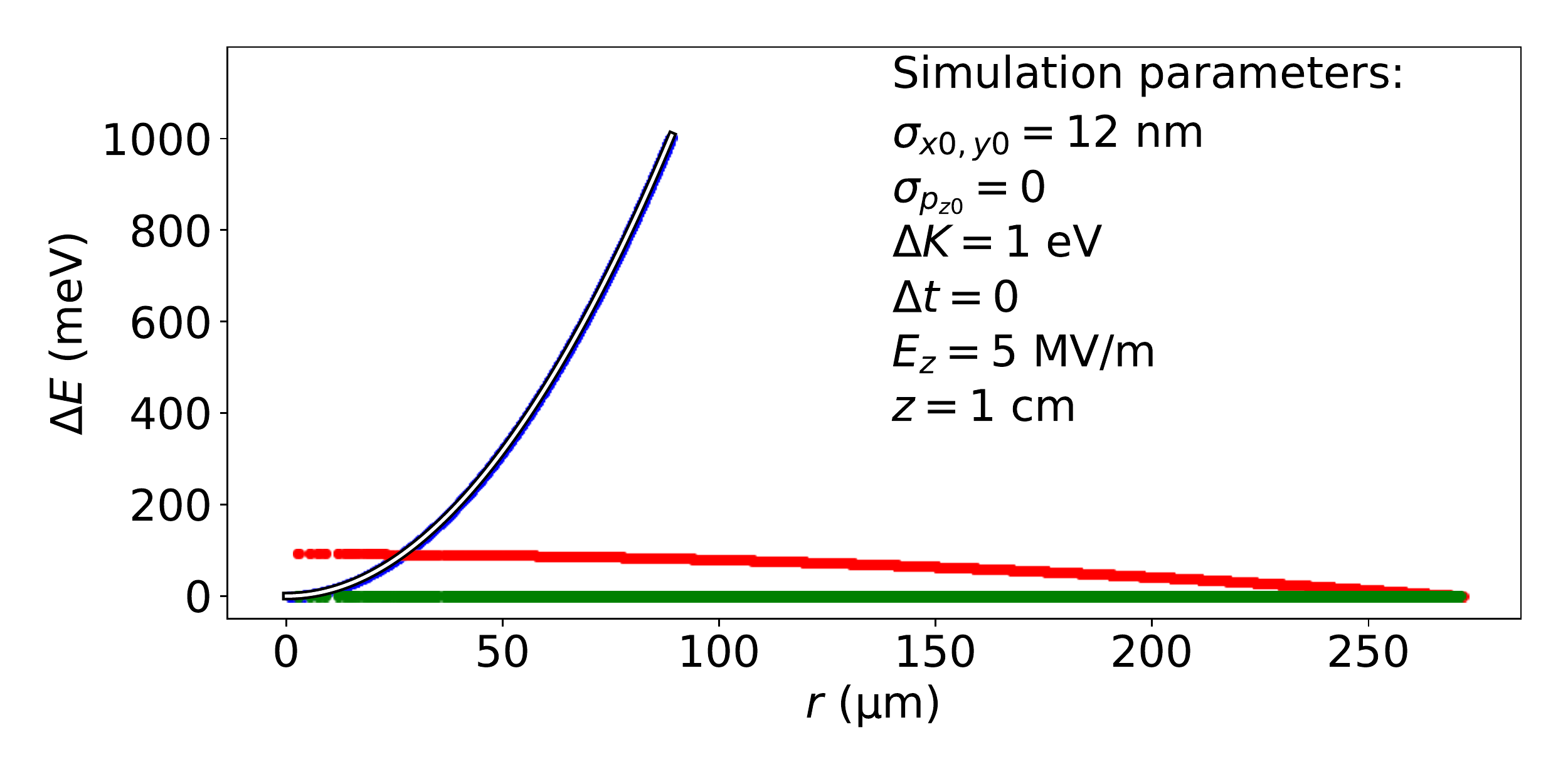}
    \caption{Simulation of a beam with initial size $\sigma_x =$ 12 nm and nonzero transverse momentum spread. Here the beam has no longitudinal momentum spread and has zero duration. Blue points show the energy-space correlation just after the gun, and red points show the result after transiting a single cavity with amplitude set by  Eq.~\eqref{rcav}. The white line is the prediction of Eq.~\eqref{K0}. Due to thick lens effects, the cavity settings predicted by Eq.~\eqref{rcav} slightly overcorrect the energy-space correlation. The green curve is the result of numerical optimization of the cavity amplitude.     }
    \label{oneDr}
\end{figure}

\begin{figure*}
    \centering
    \includegraphics[width=\linewidth]{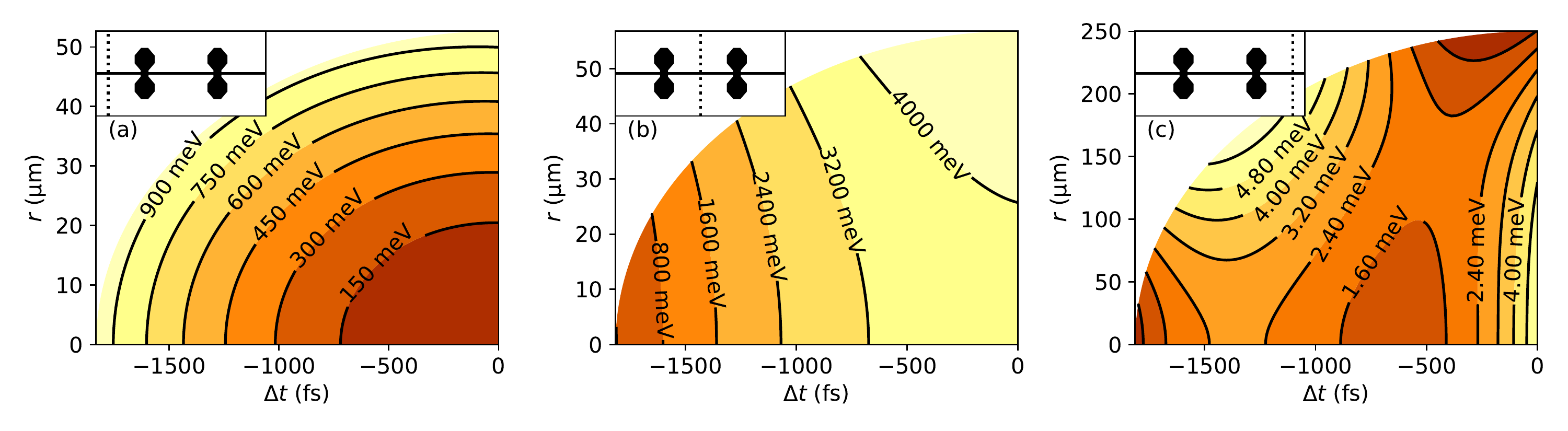}
    \caption{Simulated evolution of energy correlations on transiting two rf cavities, with an initial energy spread of $1 \ \mathrm{eV}$, vanishing initial transverse source size, and vanishing initial pulse length: (a) post gun but before the first cavity, the paraboloid distribution predicted by Eq.~\eqref{K0}; (b) between the two cavities, the first cavity imparts a hyperboloid distribution as predicted by Eq.~\eqref{cavity}, with energy increasing in time and space; (c) after the second cavity, the residual energy spread involves cubic corrections to Eq.~\eqref{K0}, shown in Eq.~\eqref{cubiccorrection}.}
    \label{finalform}
\end{figure*}

\begin{figure}
    \centering
    \includegraphics[width=\linewidth]{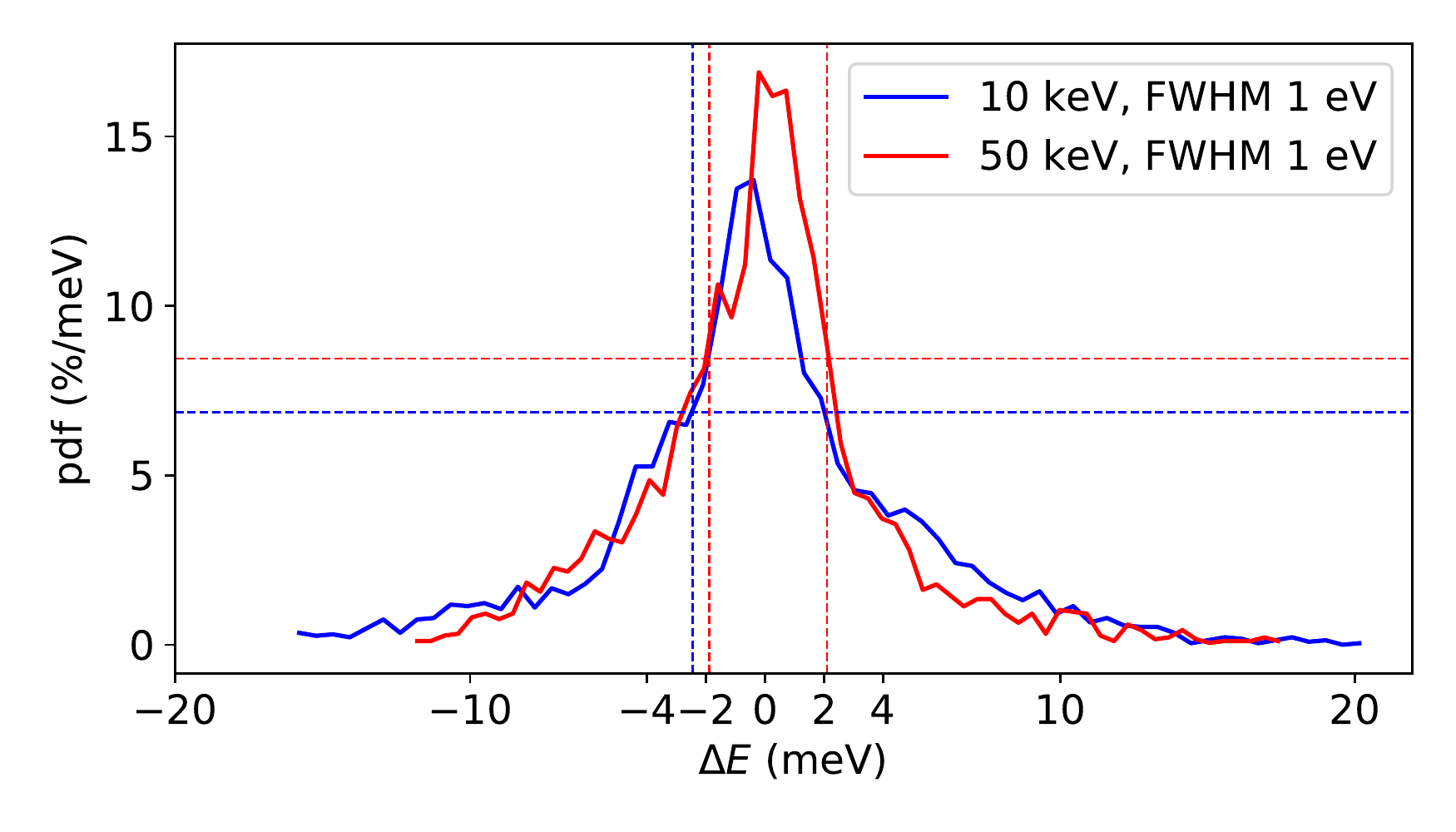}
    \caption{Results of simulating the beamline shown in Fig.~\ref{allfields}(a). Particle ensembles initially have uniform energy spreads of $1 \ \mathrm{eV}$, distributed uniformly in solid angle over the forward hemisphere. Results for two accelerating voltages are shown, $10 \ \mathrm{kV}$ (blue), $50 \ \mathrm{kV}$ (red). Dashed horizontal and vertical lines indicate the FWHM.}
    \label{nojitter}
\end{figure}

This section presents simulation results for four sets of initial beam conditions, to elucidate the beam dynamics and indicate the practical utility of our monochromator design. The first three sets are highly idealized and designed to isolate in turn the longitudinal and transverse degrees of freedom (Figs.~\ref{oneDt} and \ref{oneDr}), as well as the effects of cubic and higher-order nonlinear cavity fields on final energy spread (Fig.~\ref{finalform}). The fourth set corresponds to realistic photoemission beam size and momentum spread and demonstrates monochromation from $1 \ \mathrm{eV}$ to final energy spreads on the few meV scale (Fig.~\ref{nojitter}). The cases considered do not include the effects of timing jitter, which are accounted for in Sec.~\ref{jittersection}.

Our physics simulations are performed in General Particle Tracer (GPT) \cite{van2001general}, a Runge-Kutta particle tracking code. Simulation beamline elements and field maps are shown in Figs.~\ref{gun}, \ref{allfields}. The GPT physics model does not include the photoemission process. Instead, our initial statistical distributions of particle transverse position, time and momenta are intended to replicate the results of photoemission, as summarized in Table~\ref{inittab}. The physics of photoemission is an active area of theoretical and experimental research. We allow ourselves a margin of safety by assuming a larger initial spread of energies and emission times than has been experimentally demonstrated with cold photocathodes \cite{karkare2020ultracold}. The uncorrelated statistical uncertainty in our initial particle ensemble thus leads to poorer performance in simulations of our monochromator than would be expected with a more physically accurate treatment of photoemission.

\begin{table}
    \begin{tabular}{ c | c }
    Dynamical variable $\quad$ & $\quad$ Distribution type   \\ \hline
    Time & Uniform\\
    Energy & Uniform \\
    Angle & Uniform\\
    Position & Gaussian
    \end{tabular}
    \caption{\label{inittab}  Initial particle distributions assumed in all simulations. Measures of spread for each variable in particular simulation runs are indicated in the main text and figures.}
\end{table}

Simulating quasi-one-dimensional initial beam conditions allows for easy graphical evaluation of the accuracy of Eqs.~\eqref{tcav} and \eqref{rcav}. The first set of initial conditions is designed to isolate the longitudinal phase space, having a source with vanishing transverse size, no transverse momentum and a laser pulse length of $\Delta t_\ell = 30$ fs. Figure~\ref{oneDt}(a) shows in blue a scatter plot of this beam in energy and time as it exits the gun. The simulated gun field map is a uniform gradient with a total accelerating voltage of $50 \ \mathrm{kV}$ over a distance of $1$ cm. The results in Fig~\ref{oneDt}(a) are in close agreement with Eq.~\eqref{K0}, shown in the same figure as a white curve. 

Figure~\ref{oneDt}(a) shows in red the final distribution of the same beam in energy and time after transiting a single cavity with amplitude chosen according to Eq.~\eqref{tcav}. Because the beam is quasi-one-dimensional, the second cavity is unnecessary. Inspection of the figure shows that the cavity removes the energy spread up to a small residual linear correlation apparent in the figure at negative arrival times. Equation~\eqref{residual2} accurately predicts the coefficient of this linear correlation, shown in Fig~\ref{oneDt}(a) as a white line, confirming that the cavity restores the longitudinal emittance to its initial value.

Figure~\ref{oneDt}(b) shows results for the same quasi-one dimensional distribution as Fig.~\ref{oneDt}(a) but now accelerated in the non-uniform gun field map shown in Fig.~\ref{gun}(b). The field is nonuniform because of the electrode geometry, also shown in Fig.~\ref{gun}(b). The accelerating voltage is $50 \ \mathrm{kV}$ over a distance of $1$ cm. The field on the cathode is $1.8$ MV/m, less than half the average gradient in the gun. The prediction of Eq.~\eqref{K0} is again shown by the white curve. In evaluating Eq.~\eqref{K0}, $E^\mathrm{cat}_z$ is taken to be the field on the cathode. These results show that it is indeed the photocathode field, rather than the voltage or average gradient, which sets the final pulse length and the correlation between time of arrival and energy. The red curve shows the final energy time distribution after transiting a single cavity with amplitude chosen according to Eq.~\eqref{tcav}, again identifying the accelerating gradient with the field on the cathode.

The second set of initial conditions isolates the transverse direction. The beam at the source has nonzero size and transverse momentum spread but vanishing longitudinal momentum spread, so that all particles are emitted parallel to the photocathode surface. Figure~\ref{oneDr} shows in blue the distribution of energy against radial displacement as this beam exits the gun. The prediction of Eq.~\eqref{K0} is shown as a white curve and again agreement is close. In this case, the gun field map is a uniform accelerating gradient, which simplifies the transverse focusing in the gun. A nonuniform accelerating field results in a lens effect that spoils the agreement between the transverse distribution and Eq.~\eqref{K0}, but the qualitative picture is the same, with the energy depending quadratically on radial displacement. Figure~\ref{oneDr} shows in red the beam after transiting a single cavity with amplitude chosen according to Eq.~\eqref{rcav}. Again, because the beam is quasi-one-dimensional, the first cavity appearing in Eq.~\eqref{rcav} is unnecessary. The magnification factor $M_2$ is set equal to the ratio of beam sizes at the gun exit and cavity midplane. Inspection of the red ensemble in Fig.~\ref{oneDr} shows that a cavity with the analytically estimated settings overcorrects the energy-space correlation by 10\%. Numerically optimizing the cavity amplitude then eliminates the overcorrection, as Fig.~\ref{oneDr} shows in green.

Next, we simulate  point source initial conditions with instantaneous emission, which isolates the evolution of correlated energy spread. The results of this simulation are presented in Fig.~\ref{finalform}. The initial energy spread is 1 eV with momenta uniformly distributed in solid angle over the forward hemisphere.  The gun is modeled with the realistic fields shown in  Figure~\ref{allfields}(a). The cavity amplitude and phases are optimized numerically. Parameter values calculated from Eqs.~\eqref{tcav} and \eqref{rcav} provide good initial guesses for the optimization algorithm and consistently overestimate the optimal cavity amplitudes, which likely arise from thick lens (nonimpulsive) effects, as evidenced by inspection of individual simulated particle tracks. The sequence of transformations presented in the three panels of Fig.~\ref{finalform} confirms qualitatively the analytic approximation in Eq.~\eqref{cavity} for the work done by the cavities on the beam. Stepping through the panels, the initially paraboloid energy surface in Fig.~\ref{finalform}(a) is transformed after transit through the first cavity into a hyperboloid in Fig.~\ref{finalform}(b). Transit through the second cavity produces a surface of constant energy in Fig.~\ref{finalform}(c), up to cubic corrections of order $10^{-3}$ the initial energy spread, a scale consistent with the expression for the leading cubic correction in Eq.~\eqref{cubiccorrection}. 

Having tested the accuracy of our analytic model with idealized beam distributions, we present the results of simulations with realistic initial conditions. Figure.~\ref{gun} shows gun electrode geometry, axial field profile and the layout of the remaining simulated beamline. The simulation beamline includes exactly one focusing solenoid downstream of the gun. A minimum of two such lenses is required to image the gun exit successively at the midplanes of each cavity, the transport we assume for the sake of convenience in deriving the analytic results of Sec.~\ref{energyequalization}. Simulation results not reported here reveal that a two-solenoid design does not outperform a single-solenoid design.  The simulation cavity field map is shown in Fig.~\ref{allfields}. Figure~\ref{nojitter} shows the final energy distribution obtained from simulating a source of $12 \ \mathrm{nm}$ rms transverse radius and a uniform $1 \ \mathrm{eV}$ energy spread distributed uniformly in solid angle over the forward hemisphere. The two curves in Fig~\ref{nojitter} correspond to primary energies of $10$ and $50 \ \mathrm{keV}$. The final full width at half maximum (FWHM) for both primary energies is $4 \ \mathrm{meV}$ .

\begin{figure*}
    \centering
    \includegraphics[width=0.9\linewidth]{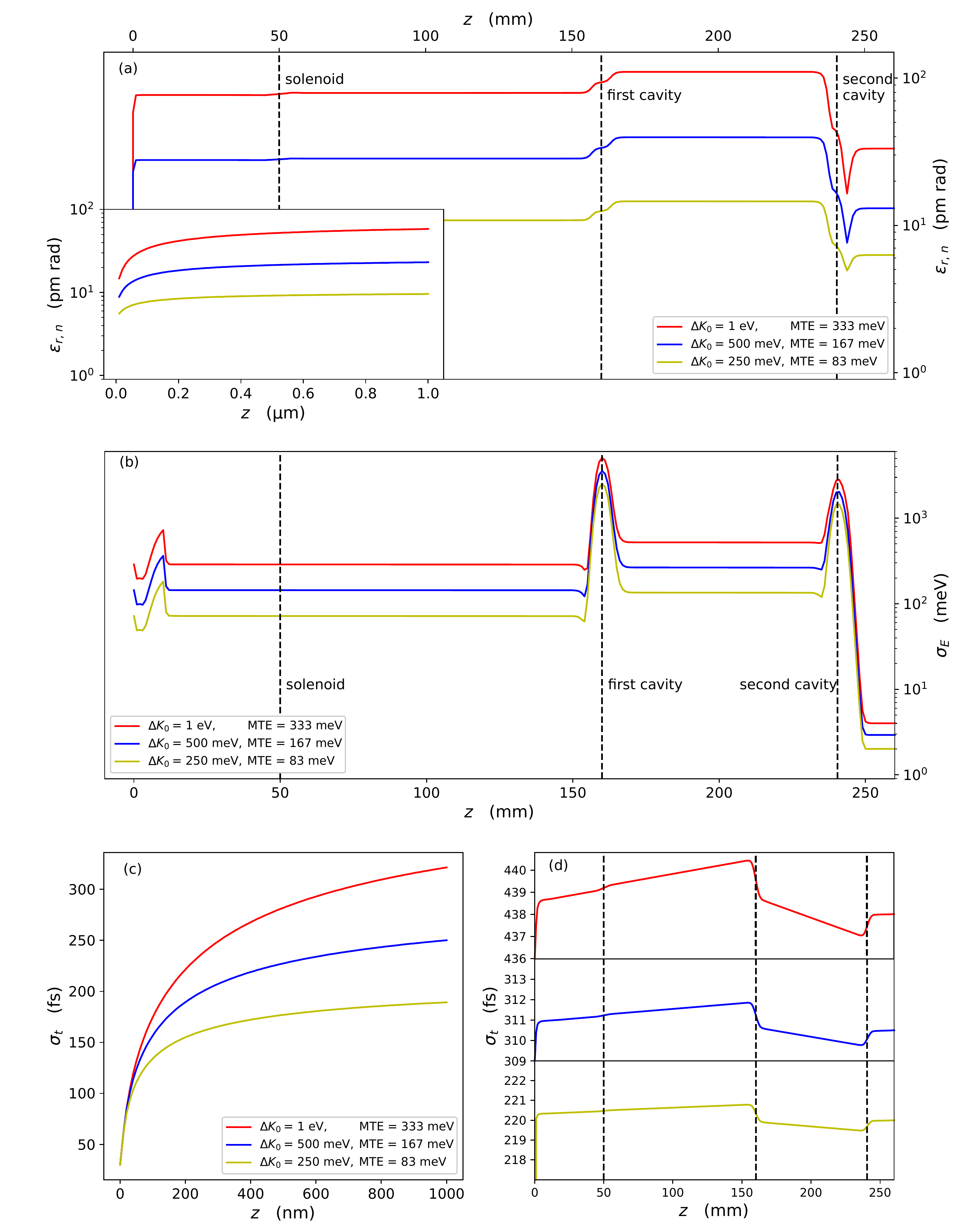}
    \caption{Simulation results showing beam sizes and emittance as a function of beamline position for three different initial energy spreads; all particle ensembles have initial transverse size of $12 \ \mathrm{nm}$ rms and momenta distributed uniformly over the forward hemisphere; with initial uniform energy spreads of $1 \ \mathrm{eV}$ (red), $500 \ \mathrm{meV}$ (blue), and $250 \ \mathrm{meV}$ (yellow). Cavity settings are chosen to achieve the smallest final energy spread. (a) Transverse normalized emittance; (b) rms energy spread; (c), (d) rms pulse duration. }
    \label{longitudinal}
\end{figure*}

The evolution of energy spread as function of position down the simulated beamline is shown in Fig.~\ref{longitudinal}(b), alongside pulse length in Fig.~\ref{longitudinal}(c) and (d). Comparing Fig.~\ref{longitudinal}(b) with Figs.~\ref{longitudinal}(c) and (d), the final value of the product $\sigma_{\cal E}\sigma_t$ is less than the initial value, in apparent contradiction with Eq.~\eqref{lemittance}. The contribution the transverse momenta make to energy spread accounts for the discrepancy, with a significant fraction of the reduction in energy spread being made possible by expanding the transverse size of the beam. The ratio of final-to-initial transverse size is order $10^3$ compared to the order 10 ratio of final-to-initial pulse lengths. A system of magnetostatic lenses can subsequently demagnify the beam without affecting energy spread and in the process further stretch the pulse length. For the initial conditions $\Delta K = 250 \ \mathrm{meV}$ shown in Fig.~\ref{longitudinal}, the final product $\sigma_{\cal E}\sigma_t$ exceeds $\hbar/2$ by only 30\%. Even as the quantum limit is approached, summary statistics calculated from classical particle tracking continue to describe the size of the beam envelope  \cite{baum2017quantum}.

\section{Jitter \label{jittersection}}

\begin{figure}
\includegraphics[width=0.5\textwidth]{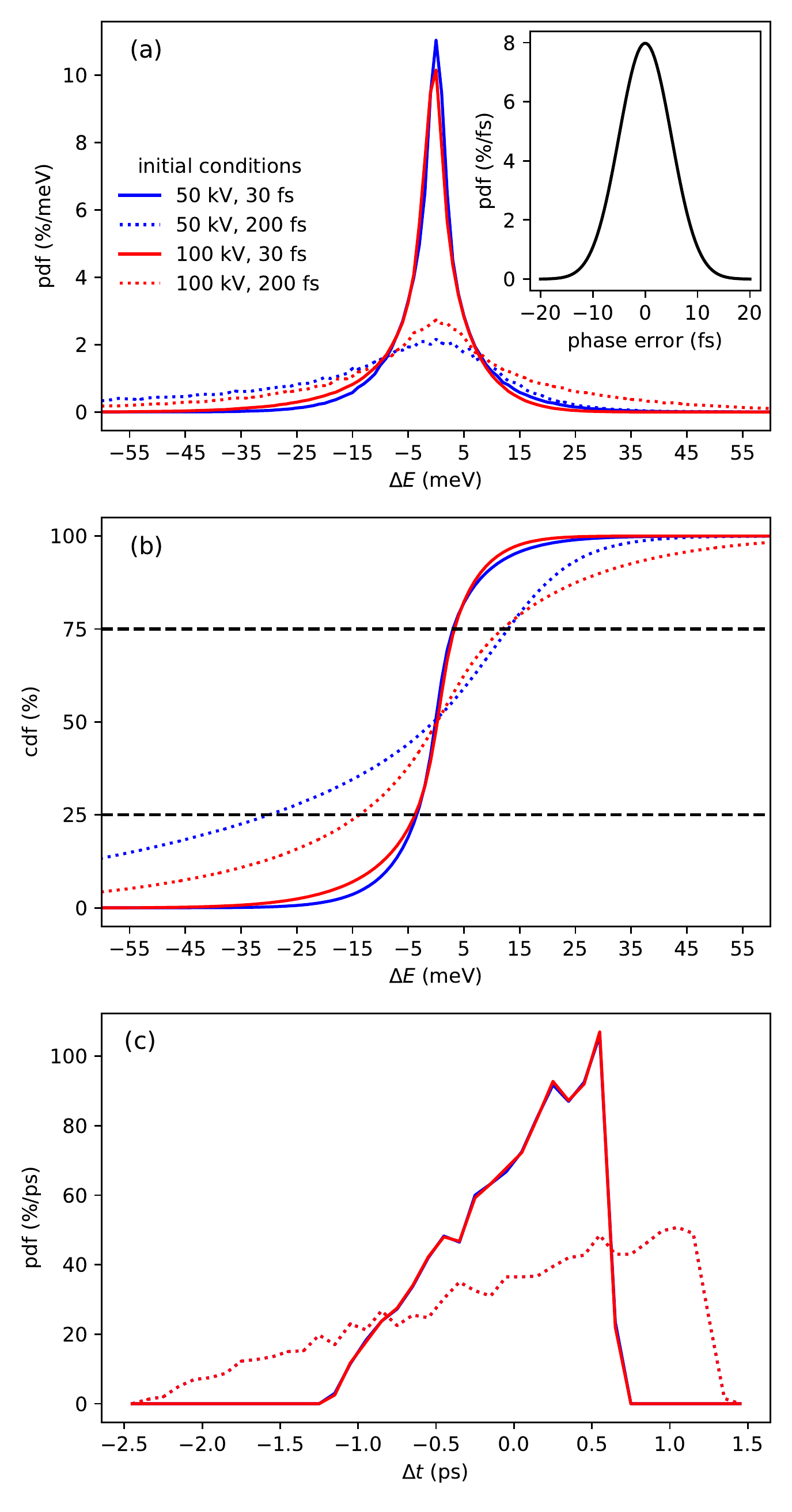}
\caption{\label{integrated}
Simulated energy distribution in the presence of white rf phase noise at the same 5 fs rms scale reported experimentally in \cite{otto2017solving}. Four sets of initial conditions are shown, with initial energy spreads of 1 eV in all. Colors indicate two sets of primary energies: 50 keV (blue), 100 keV (red). Line style indicates the initial pulse length; 30 fs (solid line), 200 fs (dotted line). (a) Probability density per meV downstream of the rf monochromator. Inset shows the distribution of the timing noise in the simulation, with the noise identically and independently distributed in each cavity. (b) Cumulative probability density with dashed horizontal lines indicating interquartile range. (c) Probability of time of arrival per picosecond relative to the mean, downstream of the gun.}
\end{figure}

Jitter in the phase of the cavities relative to the photoemitting laser pulse train contributes to the final energy spread. The study in \cite{pasmans2013microwave} finds that the scale of phase jitter is set by thermal fluctuations in the length of the cavities. The fundamental frequency of the cavity varies with changes in the cavity length, thereby shifting the phase difference between the driving oscillator and cavity response. In terms of a change in temperature $\Delta T$, the change in phase $\Delta \phi$ is \cite{pasmans2013microwave},
\begin{equation}
    \Delta \phi = -2Q\kappa_T \Delta T,
\end{equation}
where $Q$ is the unloaded quality factor of the resonator, typically $10^4$ for a normal conducting copper cavity, and $\kappa_T$ is the coefficient of thermal expansion, $1.64\times10^{-5} \ \mathrm{K}^{-1}$ for copper. At a temperature stability of $1 \ \mathrm{mK}$, the uncertainty in phase is $\Delta \phi = 3.3\times 10^{-4}$, or $17 \ \mathrm{fs}$ at $3 \ \mathrm{GHz}$. 

A phased locked loop was experimentally shown in \cite{otto2017solving} to reduce the rms phase noise of a bunching cavity to 5 fs when averaged over a 16 hour time series, and to eliminate long-term phase drift. 
 Figure~\ref{integrated} shows the simulated broadening of the zero-loss peaks (ZLP) due to phase jitter for four sets of initial conditions when the phase correction scheme of \cite{otto2017solving} is implemented. The simulations add to the phase of each cavity independent 5 fs rms Gaussian white noise. The assumption of white noise is a simplification with respect to the 1/f noise typically encountered in electron microscopy. Our white noise model therefore significantly overestimates the size of phase fluctuations on time periods of 100 ms or less that are typical for EELS acquisition. 
 
 The solid blue lines in Fig.~\ref{integrated} correspond to a primary energy of 50 keV, a 1 eV energy spread, and an initial pulse length of 30 fs. Figure ~\ref{integrated}(a) shows a FWHM of 5 meV, a broadening of 1 meV compared to the jitter-free result shown in Fig.~\ref{nojitter}. The impact of tails in the ZLP is more easily visualized in a plot of the cumulative energy distribution downstream of the monochromator, shown in Fig.~\ref{integrated}(b), which reveals that 50 \% of particles lie in a 6 meV bandwidth. Figure~\ref{integrated}(c) shows the spread of arrival times at the monochromator exit, the trade-off imposed by narrowing the energy spread. The solid red lines in Fig.~\ref{integrated} correspond to a primary energy of 100 keV, a 1 eV energy spread and an initial pulse length of 30 fs. The greater primary energy is achieved by lengthening the gun field map shown in Fig.~\ref{gun}, leaving the field strength on the cathode unchanged. Comparison of the solid blue and red lines show that the change in primary energy has little effect on the energy resolution and final pulse length.
 
 The dotted blue and red lines in Fig.~\ref{integrated} correspond to 200 fs initial pulse length and 1 eV initial energy spread. Dotted blue lines correspond to a primary energy of 50 keV and dotted red lines to 100 keV. A primary energy of 100 keV and pulse length of 200 fs is typical for ultrafast electron microscopy with field emission electron sources. The value 200 fs is representative because it approaches the bandwidth limit of Yb gain media, which are commonly chosen for high power, higher repetition rate lasers. Established techniques for increasing bandwidth allow a 200 fs pulse to be compressed to the 10 fs scale \cite{cardin2018high}.  A greater uncertainty in emission time demands a compensating reduction in the field strength on the cathode, here to 1 MV/m, performed in the simulation by scaling the dimensions of the gun-field map shown in Fig.~\ref{gun}. The dotted red line in Fig.~\ref{integrated}(a) shows a 22 meV FWHM, and the dotted red line in Fig.~\ref{integrated}(b) shows that  50\% of particles lie in a 26 meV bandwith. The spread of arrival times downstream of the gun, shown as a dotted red line in Fig.~\ref{integrated}(c), is 3 ps. The energy width figures for the 200 fs initial pulse are five times worse than for the 30 fs length. To improve the resolution of the 200 fs pulse to the same level as the 30 fs pulse would require stretching the final pulse length a further factor five, or 5\% of the 3 GHz cavity period. At a duty cycle of 5\%, the problem of optimizing cavity parameters goes beyond the quadratic approximations made in Sec.~\ref{energyequalization}, and thus we do not explore this regime in the simulation.

The time-averaged uncorrelated energy spread introduced by phase fluctuations follows by expanding the sinusoidal time dependence of the work done by the cavities,
\begin{equation}\label{jitter}
    \Delta {\cal E} = -ed_\mathrm{rf}\left(E^\mathrm{rf}_{\mathrm{max},1} + E^\mathrm{rf}_{\mathrm{max},2}\right)\left(\frac{1}{2}\Delta \phi^2 - \frac{\omega\sigma_{pz0}}{eE^\mathrm{cat}_z} \Delta \phi \right),
\end{equation}
with $\sigma_{pz0}$ the spread in initial longitudinal momenta. The factor containing $p_{z0}$ is the estimate of pulse length obtained from Eq.~\eqref{pz0}. For particles arriving earlier than average, the linear term in $\Delta \phi$ dominates because pulses are $100 \ \mathrm{fs}$ to single picoseconds long for accelerating gradients below $10 \ \mathrm{MV/m}$ at an initial energy spread of $1 \ \mathrm{eV}$ or more. However, the statistical distribution peaks at late arrival times, if the particle ensemble has an initially uniform distribution in energy. For particles near this peak, the term quadratic in $\Delta \phi$ dominates the right-hand side of Eq.~\eqref{jitter}. The takeaway from Eq.~\eqref{jitter} is thus that phase fluctuations move the tails of the energy distribution much more than they do the location of the peak, by a factor 10 to 100. 

\begin{figure*}
\includegraphics[width=1.0\textwidth]{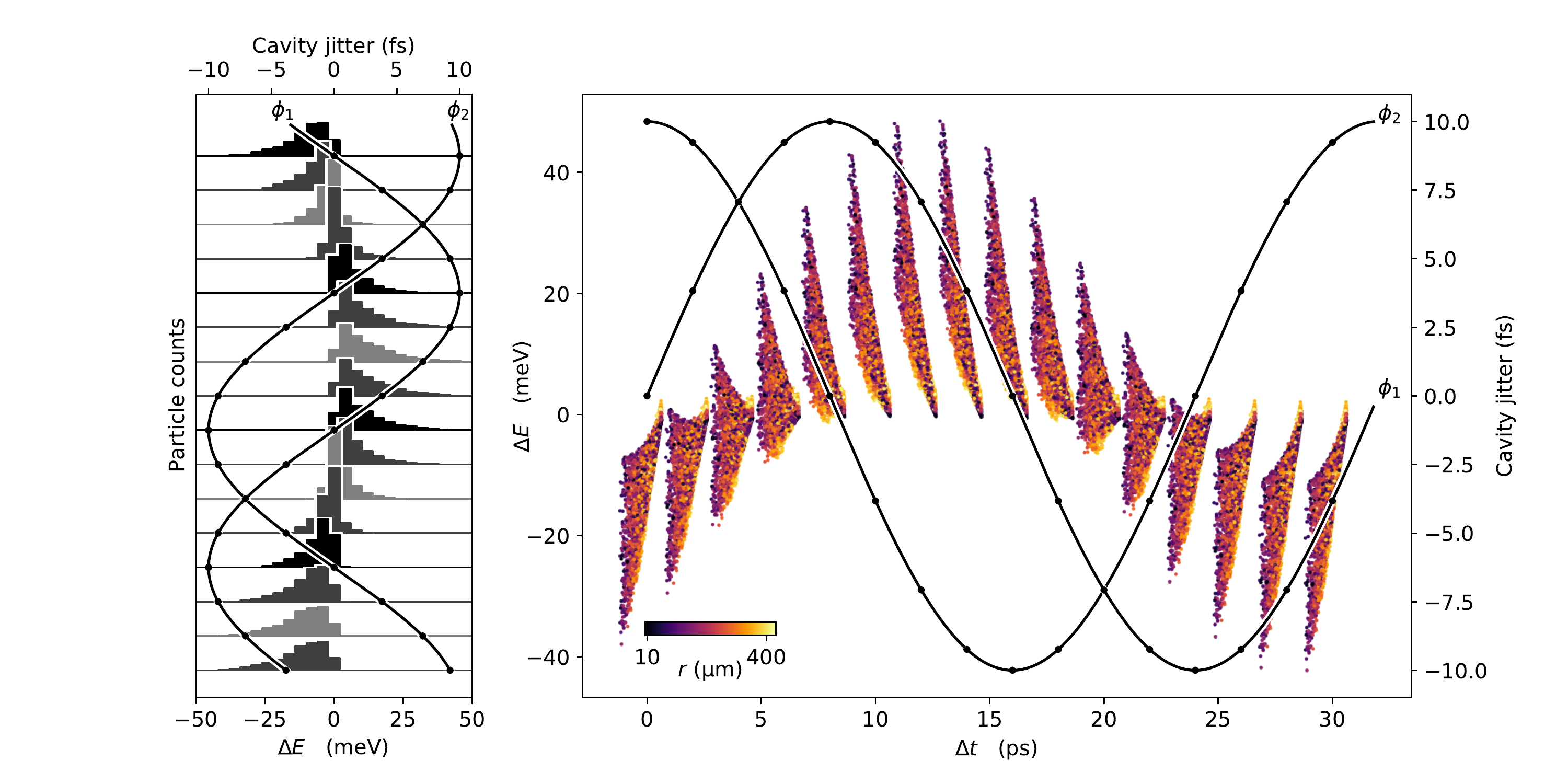}
\caption{\label{jitterhistogram} Simulations of cavity timing jitter. The initial particle ensemble has a uniform $1 \ \mathrm{eV}$ energy spread distributed uniformly in solid angle over the forward hemisphere. Plots show the ensemble after transiting two $3 \ \mathrm{GHz}$ cavities with timing offset from the optimal solution by $\phi_1, \phi_2$ respectively. Offsets add in quadrature to $10 \ \mathrm{fs}$. At left, histograms of final energy show single $\mathrm{meV}$ movement in peak location and $10 \ \mathrm{meV}$ scale movement in tails. On the , energy-time correlation are shown as a function of cavity phase, where different phases are offset on the time axis for clarity.}
\end{figure*}

At 1 mK temperature stability in the absence of rf phase correction, Eq.~\eqref{jitter} predicts a broadening near the peak of the probability distribution of less than one meV and a broadening in the tails of tens of meV. The prefactor in Eq.~\eqref{jitter} is $7 \ \mathrm{keV}$ for our simulation example of a $50 \ \mathrm{kV}$ gun with nonuniform gradient shown in Fig.~\ref{allfields}, and $1 \ \mathrm{keV}$ for the $10 \ \mathrm{kV}$ example. These estimates are confirmed in the simulation, as shown in Fig.~\ref{jitterhistogram}. The simulations cycle through a range of phase offsets $\Delta \phi_1, \Delta \phi_2$, one independent offset for each of the two cavities. For each pair of phase offsets, the right panel of Fig.~\ref{jitterhistogram} shows a scatter plot in arrival time and the final energy of particles in the beam. Comparison of the scatter plots supports our interpretation of Eq.~\eqref{jitter}, indicating in particular that the jitter-induced movement in the tails of the final energy distribution is due to early arriving particles. The left panel of Fig.~\ref{jitterhistogram} shows the corresponding sequence of final energy histograms. On inspection, the jittery peak locations remain within order $1 \ \mathrm{meV}$ the nominal peak location at $\Delta {\cal E} = 0$.

Comparatively slow fluctuations in the gun voltage are uncorrelated with differences in time of arrival between successive pulses at the relevant order of precision, according to Eq.~\eqref{time}. Energy spread due to these fluctuations is therefore not removed by our rf monochromator design. One strategy for eliminating this sub-leading source of energy spread is to implement a feedback loop, such as is included in the magnetic prism monochromator design \cite{krivanek2009high}. Our implementation would place an energy selector downstream of the cavities with a slit width chosen so that  all particles are accepted at the nominal accelerating voltage. As the accelerating voltage changes, particles clipping the slit edge trigger the feedback mechanism to change the gun high-voltage set point.

\section{Brightness conservation \label{brightness}}

To achieve maximum gains in average current, our design must omit transverse collimation. The ensemble of particles transported to the experimental target consequently includes large excursions from the optical axis, which are typically ignored in electron-microscope design. Brightness averaged over all emission angles and positions is therefore the more informative figure of merit for our unconventional beam, rather than the peak brightness more commonly encountered in a microscopy context. Our preferred figure of merit is the phase space area occupied by the beam, which is inversely proportional to the square root of the brightness. The most general measure of phase space area is {\em normalized transverse emittance}, defined as:
\begin{equation}\label{xemittance}
    \epsilon_{nx}= \frac{1}{m_e c} \sqrt{\langle x^2 \rangle \langle p_x^2 \rangle - \langle x p_x \rangle^2}.
\end{equation}
Minimum physically achievable emittance corresponds to a normalized emittance equal to half the reduced Compton wavelength of the electron. At the source, the cross term in $x$ and $p_x$ drops out of Eq.~\eqref{xemittance} and it is convenient to define a new quantity, the {\em mean transverse energy} (MTE):
\begin{equation}
    \mathrm{MTE} = \frac{\langle p_x^2\rangle}{m_e}.
\end{equation}
 For a statistical distribution that is uniform in energy and uniformly distributed in solid angle over the forward hemisphere, the MTE is equal to $2/3$ the mean energy. Letting $\sigma_{x0}$ be the rms source size, the source emittance is then,
 \begin{equation}
     \epsilon_{nx}(z=0) = \sigma_{x0}\sqrt{\frac{\mathrm{MTE}}{m_ec^2}}.
 \end{equation}

Our simulations show significant emittance degradation occurring just after emission for nanometer-sized sources with energy spreads on the 100 meV scale. Surprisingly, we find that the action of the monochromator largely undoes this emittance growth.  The emittance growth we observe can be understood as the contribution of the uniform accelerating field to the spherical and chromatic aberration of the optical column \cite{Scheinfein1993}, which amounts to a series expansion of the emittance around a vanishing solid angle. However, since our design is unlike a conventional microscope, this section explicitly derives time-of-arrival dependent expressions for the emittance growth valid at all emission angles. These expressions then predict that, up to the accuracy of the analytical model of energy-equalization presented in Sec.~\ref{energyequalization}, the final emittance after transiting the monochromator cavities is equal to the source emittance. Simulation results reported in Fig.~\ref{longitudinal} show that, beyond the rigid-beam approximation, parameters optimised for energy spread reduction overcorrect the emittance loss, imparting to the beam a correlation between time of arrival and divergence that is opposite in sign to the gun.

During acceleration of a pulsed beam, correlations evolve between time of arrival and beam divergence, so that projecting onto the transverse phase space results in brightness loss. Starting from the solution to the equations of motion in $x$ for a uniform accelerating field, the transverse emittance at proper time $\tau$ is,
\begin{equation}\label{egrowth}
    \epsilon_{n,x}^2 = \frac{\langle x_0^2\rangle\langle p_{x0}^2\rangle}{m_e^2c^2} + \frac{\langle p_{x0}^2\rangle\left\langle p_{x0}^2\tau^2\right\rangle}{m_e^4c^2} - \frac{\left\langle p_{x0}^2\tau\right\rangle^2}{m_e^4c^2}.
\end{equation}
The simplest physically plausible picture of emission from a flat cathode is that the initial momenta are uniformly distributed in solid angle over the forward hemisphere, implying a correlation between $p_{x0}$ and $p_{z0}$, and hence between $p_{x0}$ and $\tau$. These correlations are better disentangled by going over to polar coordinates, 
\begin{align}
p_{x0} &= p_0 \sin\phi\cos\theta, \\
p_{z0} &= p_0\cos\theta,
\end{align} 
and assuming that the momentum magnitude $p$ is uncorrelated with the emission angles $\phi, \theta$. Letting the probability of emission be uniform in azimuth $\phi$, expanding $\tau$ to first order in $\Delta t$ per Eq.~\eqref{tau}, and substituting the expression for $\Delta t$ in terms of $\Delta p_{z0}$ given by Eq.~\eqref{pz0}, the result is that Eq.~\eqref{egrowth} becomes
\begin{equation}\label{esource}
    \epsilon_{n,x}^2 = \frac{\langle \beta_{0}^2\rangle}{3}\left(\langle x_0^2\rangle +
    \sigma^2_{x0*}\right),
\end{equation}
with
\begin{align}\label{size}
    \sigma^2_{x0*} = 
    \frac{m_e^2c^4}{e^2E_z^2}\left(\frac{1}{15}\left\langle \beta_{0}^4\right\rangle-\frac{3}{64}\frac{\left\langle \beta_{0}^3\right\rangle^2}{\left\langle \beta_{0}^2\right\rangle}\right).
\end{align}
where, again, $E_z$ is the longitudinal field in the gun. The term ${\sigma_{x0*}}^2$ is a new, critical feature of near-cathode dynamics for nanometer-scale photoemission sources: an effective source size growth arising from photoemission momentum spread. The scaling of this effective source size with mean transverse energy has practical importance for active research toward higher brightness photocathodes. Assuming a uniform distribution in energy at the source, Eq.~\eqref{size} simplifies to,
\begin{equation}
    \sigma_{x0*} = 0.363\times\frac{\langle \mathrm{MTE} \rangle }{|eE_z|}.
\end{equation}
Thus, for nano-scale sources, photocathode emittance goes like $\mathrm{MTE}^{3/2}$ and not the expected $\mathrm{MTE}^{1/2}$, enhancing the potential impact of new low MTE materials on the future performance of photoemitters \cite{musumeci2018advances}.

\begin{figure}
    \centering
    \includegraphics[width=\linewidth]{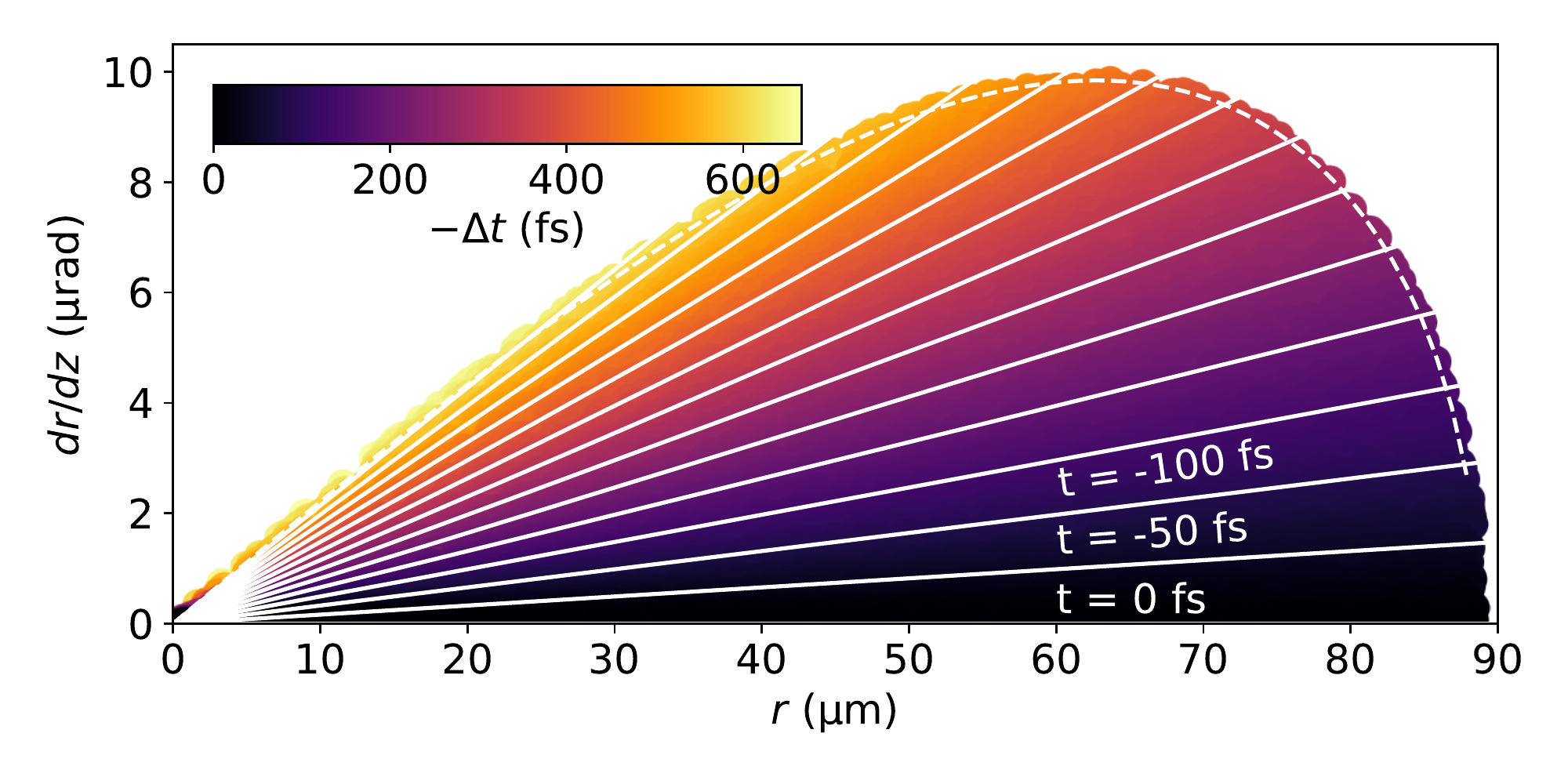}
    \caption{Correlations between particle divergence and time of arrival in particle tracking simulations of a uniform accelerating gradient of $ 5 \  \mathrm{MV/m}$. The initial conditions of the particle distribution are a 1 eV uniform energy spread distributed uniformly in solid angle over the forward hemisphere, and vanishing transverse size and pulse length. The time-dependent divergence predicted by Eq.~\eqref{fgun} are shown by the white lines at constant time increments of $50 \ \mathrm{fs}$.}
    \label{slice}
\end{figure}

To see that the emittance growth expressed in Eq.~\eqref{esource} can be undone by a time-dependent lens, we invert the solution to the equations of motion in $x$ and expand $\tau$ to obtain the following expression for the particle angle $dx/dz$ as a function of $\Delta t$,
\begin{equation}\label{angle}
    \frac{p_x}{p_z} = \frac{m_e(x-x_0)}{p_{z*}\tau_*} - \frac{m_e(x-x_0)}{p_{z*}\tau_*^2}\Delta t.
\end{equation}
The term in Eq.~\eqref{angle} proportional to $\Delta t$ is the same as the expression for the change in angle imparted by an ideal thin lens with time-dependent focal length,
\begin{align}\label{fgun}
    f^{-1}_\mathrm{gun} &=  \frac{(e E^\mathrm{cat}_z)^2\Delta t}{\gamma_*\beta_* m_e^2c^3{\cal A}^2}.
\end{align}
Equation \eqref{fgun} is not meant to be an explanation of the cause of the emittance growth in the gun. Instead, Eq.~\eqref{fgun} makes the linear correlation with time of arrival explicit in a manner that serves to explain how the time-dependent focusing power of an rf cavity is able to counteract the emittance growth in the gun.

Simulation results verifying Eq.~\eqref{fgun} are shown in Fig.~\ref{slice}.  To produce the plot, the beam divergence of the time-independent term in Eq.~\eqref{angle} is subtracted. Figure~\ref{slice} overlays the divergence predicted by Eq.~\eqref{fgun} as white lines at fixed increments in time of arrival. A color scale indicates the simulation time of arrival. The curved beam envelope, indicated by a dashed white line in Fig.~\ref{slice}, is derivable from Equation \eqref{fgun} by expressing divergence as a function of initial kinetic energy:
\begin{equation}\label{envelope}
    \frac{p_x}{p_z}\left(K, \ x\right) = \frac{-eE^\mathrm{cat}_zx}{\gamma_*\beta_*{\cal A}^2}
    \sqrt{\frac{2K}{m_e^3c^6} - \frac{e^2{E^\mathrm{cat}_z}^2x^2}{m_e^4c^8{\cal A}^2}}.
\end{equation}
The envelope is found by holding $K$ at a constant value equal to the maximum initial kinetic in the particle ensemble: $1 \ \mathrm{eV}$ for Fig.~\ref{slice}. Comparing Eq.~\eqref{envelope} and Eq.~\eqref{fgun} clarifies an unexpected advantage of time-dependent focusing fields: explicit control in the time domain makes linear an effect that appears nonlinear in the transverse phase space.

The transverse focusing power of a cavity, calculated by making the same thin-lens approximation assumed in the derivation of Eq.~\eqref{cavity}, is \cite{pasmans2013microwave},
\begin{equation}\label{fcav}
    f^{-1}_\mathrm{1,2} =  \mp\frac{eE^\mathrm{rf}_{\mathrm{max},1,2}d_\mathrm{rf}}{2\gamma_*^3\beta_*^3m_ec^3}\omega^2\Delta t,
\end{equation}
where the cavity phase is taken to maximize (in magnitude) the reference particle.  

If the gun exit is successively imaged at the midplanes of the two cavities, with magnification factors $M_1$ and $M_2$, then the condition for cancellation of the emittance growth is,
\begin{equation}\label{cond}
    \frac{1}{f_\mathrm{gun}} + \frac{M_1^2}{f_1} + \frac{M_2^2}{f_2} = 0.
\end{equation}
The derivation of Eq.~\eqref{cond} assumes the pulse length remains constant after exiting the gun, a good approximation for the compact beamline shown in Fig.~\ref{allfields}(a).
Substituting the expressions in Eqs.~\eqref{fgun} and \eqref{fcav} into into Eq.~\eqref{cond} gives a third equation on the system of monochromator parameters $E^\mathrm{rf}_{\mathrm{max},1}, M_1, E^\mathrm{rf}_{\mathrm{max},2}, M_2$,
\begin{align}\label{emitcond}
    ed_\mathrm{rf}E^\mathrm{rf}_{\mathrm{max},1}M_1^2  - ed_\mathrm{rf}E^\mathrm{rf}_{\mathrm{max},2}M_2^2 =&  \frac{(e E^\mathrm{cat}_z\lambda)^2}{2\pi^2}\frac{\gamma_*^2\beta_*^2}{m_ec^2{\cal A}^2},
\end{align}
where ${\cal A}$ is again the aspect ratio defined in Eq.~\eqref{aspect}. Equation~\eqref{emitcond} is equivalent to the equation obtained by setting the energy spread to zero in Eq.~\eqref{rcav}.

Simulation results presented in Fig.~\ref{longitudinal} show that cavity parameters optimal for reducing energy spread actually overcorrect the brightness loss in the gun. Simulated particle tracks show that this overcorrection is due to particles not obeying the rigid-beam approximation made in deriving Eq.~\eqref{rcav}. Thus, there is a trade-off between energy spread reduction and brightness conservation in a scheme involving only two cavities. Simulations show that, at the cost of reintroducing energy spread at the $10^{-1}$ level compared to the source, reducing the second cavity amplitude does perfectly restore the source emittance. Figure~\ref{longitudinal} shows simulation results for three values of the source energy spread: uniform distributions over $1 \ \mathrm{eV}, \ 500 \ \mathrm{mev}$ and $100 \ \mathrm{meV}$. The trend in energy reveals that at $250 \ \mathrm{meV}$ and below, the excess final emittance at the minimum achievable final energy spread exceeds the source emittance by less than 30\%. 

\section{Discussion and conclusion}

This paper has shown in simulation the feasibility of a high-energy resolution, pulsed electron source design. The design utilizes radiofrequency accelerating cavities to top-off the primary energy delivered by a dc gun. Relying on correlations between energy on the one hand and time and position of arrival at the cavities on the other, the additional cavity acceleration equalizes energies in the particle ensemble. An initial energy spread of $1 \ \mathrm{eV}$ is reduced in simulation to $4 \ \mathrm{meV}$ at primary energies of $10 \ \mathrm{kV}$ and $50 \ \mathrm{kV}$ assuming arbitrary precision in cavity timing. Simulating the effect of cavity jitter at $5 \ \mathrm{fs}$ timing precision shows the peak of the energy distribution broadening on the single $\mathrm{meV}$ scale. The efficacy of the design is explicable by simple yet accurate analytic expressions, strongly suggesting that our results are not sensitive to peculiar choices of simulation parameters and that the scheme could be realized experimentally with presently available technology. 

A near at hand application of our design is to install two rf cavities in an existing ultrafast electron transmission microscope. Existing UEM instruments photoemit from nanometer-scale tips, where significant field enhancement occurs. Typical gradients are $1 \ \mathrm{GeV/m}$ at nanometer distances.  The importance of field enhancement for our scheme is that the rapid acceleration of particles escaping the field-enhanced region implies a shorter final pulse length and hence a larger final energy spread, following Eq.~\eqref{lemittance}. Strategies to compensate for this unwanted side effect of field enhancement include: reducing the extraction voltage, increasing the tip radius, and introducing an extended drift space immediately following extraction.

Our results readily extend to monochromation at primary energies of hundreds of keV. At higher primary energies, the rigid-beam approximation made in our analytic treatment of the cavities becomes more accurate. Conversely, the rigid beam approximation breaks down as the primary energy approaches zero. The error in the approximation becomes significant when the transit time through a single cavity is comparable to the cavity period. For a $3$ GHz cavity with a 1 cm gap, the ratio of transit time to rf period is unity at a primary energy of $2.6$ keV. Exploring cavity monochromation at primary energies below $3$ keV thus requires a nonimpulsive treatment of the work done by the cavities. 

This paper also investigated, analytically and numerically, the effect of the rf cavities on transverse brightness. Fortuitously, time-dependent cavity lensing undoes brightness loss in the gun. Effects apparent in simulation that go beyond the impulsive approximation made in our analytic treatment result in the cavities overcorrecting the brightness loss. Simulations show that the overcorrection becomes negligible at source energy spreads of $250 \ \mathrm{meV}$ and below, a finding that underscores the importance for future electron-beam technologies of ongoing research into photocathode materials with low intrinsic energy spreads \cite{musumeci2018advances}.

\section{Acknowledgements} We thank J. P. Gonzalez Aguilera for help with particle tracking simulations. This research was supported by the Center for Bright Beams, U.S. National Science Foundation Grant No. PHY-1549132. 

\appendix
\section{Cavity field expansion}\label{cavexp}

The derivation of the field expansion shown in Eq.~\eqref{fields} proceeds by the method of Frobenius. The assumption of harmonic time dependence simplifies the wave equation to the condition that,
\begin{equation}\label{Helmholtz}
    \left( \nabla^2 + \frac{\omega^2}{c^2}\right) \sum^\infty_{n=0}\sum_{m=0}^\infty\sum_{q=0}^\infty a_{nmq}r^n \frac{\omega^m}{c^m} \frac{d^q}{d z^q}E_{0,z}^\mathrm{rf}(z) = 0,
\end{equation}
in the unknown coefficients $a_{nmq}$. The special case in which $\omega=0$ is commonly treated in textbooks on electron optics; see, e.g., \cite{zuo2017advanced}. Computing the action of the differential operator term by term, Eq.~\eqref{Helmholtz} leads to,
\begin{align}\label{HelmholtzAction}
    \sum^\infty_{n=0}\sum_{m=0}^\infty\sum_{q=0}^\infty a_{nmq}\left( n^2 r^{n-2} \frac{\omega^{m}}{c^m} \frac{d^q}{d z^q} \right. \notag \\
    \left. + r^n \frac{\omega^{m+2}}{c^{m+2}} \frac{d^q}{d z^q}  +  r^n \frac{\omega^{m}}{c^{m}} \frac{d^{q+2}}{d z^{q+2}} \right)E_{0,z}^\mathrm{rf}(z) = 0.
\end{align}
Since Eq.~\eqref{HelmholtzAction} must hold for arbitrary $\omega$ and $E^\mathrm{ref}_{0,z}(z)$, the left-hand side vanishes order by order in $r, \omega$ and $d/dz$. The implied recursion relation that the $a_{nmq}$ satisfy is therefore,
\begin{equation}\label{recursion}
    n^2a_{n, m, q} + a_{n-2, m-2, q} + a_{n-2,m,q-2} = 0,
\end{equation}
subject to the boundary conditions that $a_{0,0,0} = 1$ and that the coefficients for negative powers of $r$ vanish. These boundary conditions are necessary and jointly sufficient for the field at $r=0$ to coincide with $E_{0,z}^\mathrm{rf}(z)$. Let us state and then verify that the solution to Eq.~\eqref{recursion} is,
\begin{equation}\label{recsolve}
    a_{2n, 2m, 2q} = \left(\frac{-1}{4}\right)^n \frac{1}{n!m!q!} \quad \mathrm{if} \ n = m + q,
\end{equation}
with all remaining coefficients vanishing.
To verify, substituting the right-hand side of Eq.~\eqref{recsolve} into the left-hand side of Eq.~\eqref{recursion} gives,
\begin{align}\label{working}
\left(\frac{-1}{4}\right)^n \frac{1}{n!m!q!}\left[
4n^2 - 4nm -4nq\right] = 0.
\end{align}
The second and third terms inside the square brackets on the left hand side of Eq.~\eqref{working} follow from the property of factorials that $(n-1)! = n!/n$. Equation~\eqref{working} holds if and only if $n = m + q$; hence, we have solved the recursion relation expressed in Eq.~\eqref{recursion}. Using the constraint $n=m+q$, the solution can be simplified to
\begin{equation}
    a_{2n, 2k, 2(n-k)} = \left(\frac{-1}{4}\right)^n \frac{1}{(n!)^2}{n \choose k}.
\end{equation}
The appearance of the binomial coefficient in the simplified solution then implies that the contributions from $\omega$ and $d/dz$ sum to give
\begin{align}\label{HelmholtzFinal}
\sum^\infty_{n=0}\sum_{m=0}^\infty\sum_{q=0}^\infty a_{nmq}r^n \frac{\omega^m}{c^m} \frac{d^q}{d z^q}\notag \\ 
=\sum^\infty_{n=0} \left(\frac{-1}{4}\right)^n \frac{r^{2n}}{(n!)^2}\sum_{k=0}^n {n \choose k}  \frac{\omega^{2k}}{c^{2k}} \frac{d^{2(n-k)}}{d z^{2(n-k)}} \\
= \sum^\infty_{k=0} \left(\frac{-1}{4}\right)^k\frac{r^{2k}}{(k!)^2}\left(\frac{d^2}{dz^2} + \frac{\omega^2}{c^2} \right)^k.
\end{align}
The coefficients $(-1/4)^k /(k!)^2$ appearing on the right-hand side of Eq.~\eqref{HelmholtzFinal} are the power series expansion of the zeroth-order Bessel function of the first kind, from which Eq.~\eqref{fields} in the main text immediately follows.

\section{Integration by parts}\label{intparts}
To perform the integration by parts, we need to derive an intermediate result, namely,
\begin{equation}\label{lemma}
    \int_{-\infty}^\infty \frac{d^{2n} f(x)}{dx^{2n}}\cos(kx) dx = 
    (-1)^n k^{2n}\int_{-\infty}^\infty f(x)\cos(kx)dx,
\end{equation}
for arbitrary $f(x)$ satisfying, for all $n$,
\begin{equation}\label{fboundary}
    \lim_{x\rightarrow\pm\infty} \frac{d^nf}{dx^n}  = 0.
\end{equation}
The derivation is by mathematical induction. The base case, $n=0$, is trivially true. Now assume that Eq.~\eqref{lemma} holds for arbitrary $n$. It then follows, integrating by parts twice, that Eq.~\eqref{lemma} also holds for $n+1:$
\begin{align}
&\int_{-\infty}^\infty \frac{d^{2n+2} f(x)}{dx^{2n+2}}\cos(kx) dx \notag \\
    =&\left[\frac{d^{2n+1} f(x)}{dx^{2n+1}}\cos(kx) \right]_{-\infty}^\infty + k\int_{-\infty}^\infty \frac{d^{2n+1} f(x)}{dx^{2n+1}}\sin(kx) dx \\
    =& k\int_{-\infty}^\infty \frac{d^{2n+1} f(x)}{dx^{2n+1}}\sin(kx) dx \\
    =& \left[\frac{d^{2n} f(x)}{dx^{2n}}\sin(kx) \right]_{-\infty}^\infty - k^2\int_{-\infty}^\infty \frac{d^{2n} f(x)}{dx^{2n}}\cos(kx) dx, \label{inductive}
\end{align}
where the terms in square brackets vanish, following Eq.~\eqref{fboundary}. Applying the inductive hypothesis to the integral on the right-hand side of Eq.~\eqref{inductive}, yields
\begin{align}
    \int_{-\infty}^\infty \frac{d^{2n+2} f(x)}{dx^{2n+2}}\cos(kx) dx \notag \\=
    (-1)^{n+1} k^{2n+2}\int_{-\infty}^\infty f(x)\cos(kx)dx,
\end{align}
completing the demonstration of Eq.~\eqref{lemma}.

Making use of Eq.~\eqref{lemma} to more perform explicitly the calculation in the main text:
\begin{align}
W(r,t) = -\int_{-\infty}^{\infty} e{\bf E} \cdot d {\bf s} \approx -\int_{-\infty}^{\infty} eE_z(t(z), z, r) dz \\
= -e\sum_{n=0}^\infty \sum_{k=0}^\infty  
\left(\frac{-1}{4}\right)^n \frac{r^{2n}}{(n!)^2}\frac{\omega^{2(n-k)}}{c^{2(n-k)}} 
{n \choose k}\notag \\
\times \int_{-\infty}^\infty \frac{d^{2k}}{dz^{2k}} E_{z,0}^\mathrm{rf}(z) \cos\left(
\frac{2\pi}{\beta} \frac{z}{\lambda} + \omega\Delta t + \phi_0\right)dz, \label{notsimple}
\end{align}
where $\Delta t$ is the difference in times of arrival of the given particle and the reference particle. The cosine function expands as,
\begin{align}
 \cos\left(
\frac{2\pi}{\beta} \frac{z}{\lambda} + \omega\Delta t + \phi_0\right) =
    \cos\left(
\frac{2\pi}{\beta} \frac{z}{\lambda}\right)\cos\left(\omega\Delta t + \phi_0\right) \notag \\ 
-     \sin\left(
\frac{2\pi}{\beta} \frac{z}{\lambda}\right)\sin\left(\omega\Delta t + \phi_0\right).
\end{align}
The integral over the factor in the sine of $z$ vanishes because sine is an odd function. Hence, the expression on the right-hand side of Eq.~\eqref{notsimple} further simplifies to,
\begin{align}
W(r,t) =  -e\cos(\omega\Delta t + \phi_0)\int_{-\infty}^\infty E_{z,0}^\mathrm{rf}(z) \cos\left(
\frac{2\pi}{\beta} \frac{z}{\lambda} \right)dz \notag \\
\times \sum_{n=0}^\infty \sum_{k=0}^\infty  
\left(\frac{-1}{4}\right)^n \frac{r^{2n}}{(n!)^2}\frac{\omega^{2(n-k)}}{c^{2(n-k)}} 
{n \choose k}\left( \frac{2\pi i}{\beta \lambda} \right)^{2k}. \label{byparts}
\end{align}
The imaginary unit appearing in Eq.~\eqref{byparts} simplifies the prefactor $(-1)^k$, which the integration by parts contributed. Summing over $k$,
\begin{align}
   \sum_{k=0}^\infty  \frac{\omega^{2(n-k)}}{c^{2(n-k)}} 
{n \choose k}\left( \frac{2\pi i}{\beta \lambda} \right)^{2k} \notag \\ =
   \sum_{k=0}^\infty  \frac{\omega^{2n}}{c^{2n}}
{n \choose k}\left( \frac{i}{\beta} \right)^{2k} \\
=    \frac{\omega^{2n}}{c^{2n}} 
\left(1 - \beta^{-2} \right)^n \\ = \left(\frac{2\pi i}{\beta\gamma \lambda} \right)^{2n}.
\end{align}
% Produces the bibliography via BibTeX.
The remaining sum over $n$ in Eq.~\eqref{byparts} is the power series expansion of the zeroth-order Bessel function of the first kind; hence,
\begin{align}
    W(r,t) = -e\cos(\omega\Delta t + \phi_0)\int_{-\infty}^\infty E_{z,0}^\mathrm{rf}(z) \cos\left(
\frac{2\pi}{\beta} \frac{z}{\lambda} \right)dz \notag \\
\times \sum_{n=0}^\infty  
\left(\frac{-1}{4}\right)^n \frac{r^{2n}}{(n!)^2} \left(\frac{2\pi i}{\beta\gamma \lambda} \right)^{2n} \\
= -J_0 \left(\frac{2\pi i}{\beta \gamma \lambda} \right)\cos(\omega\Delta t + \phi_0) \notag \\ \times e\int_{-\infty}^\infty E_{z,0}^\mathrm{rf}(z) \cos\left(
\frac{2\pi}{\beta} \frac{z}{\lambda} \right)dz,
\end{align}
completing the calculation.


\begin{thebibliography}{53}%
\makeatletter
\providecommand \@ifxundefined [1]{%
 \@ifx{#1\undefined}
}%
\providecommand \@ifnum [1]{%
 \ifnum #1\expandafter \@firstoftwo
 \else \expandafter \@secondoftwo
 \fi
}%
\providecommand \@ifx [1]{%
 \ifx #1\expandafter \@firstoftwo
 \else \expandafter \@secondoftwo
 \fi
}%
\providecommand \natexlab [1]{#1}%
\providecommand \enquote  [1]{``#1''}%
\providecommand \bibnamefont  [1]{#1}%
\providecommand \bibfnamefont [1]{#1}%
\providecommand \citenamefont [1]{#1}%
\providecommand \href@noop [0]{\@secondoftwo}%
\providecommand \href [0]{\begingroup \@sanitize@url \@href}%
\providecommand \@href[1]{\@@startlink{#1}\@@href}%
\providecommand \@@href[1]{\endgroup#1\@@endlink}%
\providecommand \@sanitize@url [0]{\catcode `\\12\catcode `\$12\catcode
  `\&12\catcode `\#12\catcode `\^12\catcode `\_12\catcode `\%12\relax}%
\providecommand \@@startlink[1]{}%
\providecommand \@@endlink[0]{}%
\providecommand \url  [0]{\begingroup\@sanitize@url \@url }%
\providecommand \@url [1]{\endgroup\@href {#1}{\urlprefix }}%
\providecommand \urlprefix  [0]{URL }%
\providecommand \Eprint [0]{\href }%
\providecommand \doibase [0]{https://doi.org/}%
\providecommand \selectlanguage [0]{\@gobble}%
\providecommand \bibinfo  [0]{\@secondoftwo}%
\providecommand \bibfield  [0]{\@secondoftwo}%
\providecommand \translation [1]{[#1]}%
\providecommand \BibitemOpen [0]{}%
\providecommand \bibitemStop [0]{}%
\providecommand \bibitemNoStop [0]{.\EOS\space}%
\providecommand \EOS [0]{\spacefactor3000\relax}%
\providecommand \BibitemShut  [1]{\csname bibitem#1\endcsname}%
\let\auto@bib@innerbib\@empty
%</preamble>
\bibitem [{\citenamefont {Hawkes}(2009)}]{Hawkes2009a}%
  \BibitemOpen
  \bibfield  {author} {\bibinfo {author} {\bibfnamefont {PW}~\bibnamefont
  {Hawkes}},\ }\bibfield  {title} {\bibinfo {title} {Aberration correction past
  and present},\ }\href@noop {} {\bibfield  {journal} {\bibinfo  {journal}
  {Philosophical Transactions of the Royal Society A: Mathematical, Physical
  and Engineering Sciences}\ }\textbf {\bibinfo {volume} {367}},\ \bibinfo
  {pages} {3637--3664} (\bibinfo {year} {2009})}\BibitemShut {NoStop}%
\bibitem [{\citenamefont {Scherzer}(1949)}]{scherzer1949theoretical}%
  \BibitemOpen
  \bibfield  {author} {\bibinfo {author} {\bibfnamefont {O}~\bibnamefont
  {Scherzer}},\ }\bibfield  {title} {\bibinfo {title} {The theoretical
  resolution limit of the electron microscope},\ }\href@noop {} {\bibfield
  {journal} {\bibinfo  {journal} {Journal of Applied Physics}\ }\textbf
  {\bibinfo {volume} {20}},\ \bibinfo {pages} {20--29} (\bibinfo {year}
  {1949})}\BibitemShut {NoStop}%
\bibitem [{\citenamefont {Sch{\"o}nhense}\ and\ \citenamefont
  {Spiecker}(2002)}]{schonhense2002correction}%
  \BibitemOpen
  \bibfield  {author} {\bibinfo {author} {\bibfnamefont {G}~\bibnamefont
  {Sch{\"o}nhense}}and\ \bibinfo {author} {\bibfnamefont {H}~\bibnamefont
  {Spiecker}},\ }\bibfield  {title} {\bibinfo {title} {Correction of chromatic
  and spherical aberration in electron microscopy utilizing the time structure
  of pulsed excitation sources},\ }\href@noop {} {\bibfield  {journal}
  {\bibinfo  {journal} {Journal of Vacuum Science \& Technology B:
  Microelectronics and Nanometer Structures Processing, Measurement, and
  Phenomena}\ }\textbf {\bibinfo {volume} {20}},\ \bibinfo {pages} {2526--2534}
  (\bibinfo {year} {2002})}\BibitemShut {NoStop}%
\bibitem [{\citenamefont {Khursheed}(2005)}]{Khursheed2005}%
  \BibitemOpen
  \bibfield  {author} {\bibinfo {author} {\bibfnamefont {Anjam}\ \bibnamefont
  {Khursheed}},\ }\bibfield  {title} {\bibinfo {title} {Dynamic chromatic
  aberration correction in low energy electron microscopes},\ }\href@noop {}
  {\bibfield  {journal} {\bibinfo  {journal} {Journal of Vacuum Science \&
  Technology B: Microelectronics and Nanometer Structures Processing,
  Measurement, and Phenomena}\ }\textbf {\bibinfo {volume} {23}},\ \bibinfo
  {pages} {2749--2753} (\bibinfo {year} {2005})}\BibitemShut {NoStop}%
\bibitem [{\citenamefont {Reijnders}\ \emph {et~al.}(2010)\citenamefont
  {Reijnders}, \citenamefont {Debernardi}, \citenamefont {van~der Geer},
  \citenamefont {Mutsaers}, \citenamefont {Vredenbregt},\ and\ \citenamefont
  {Luiten}}]{Reijnders2010}%
  \BibitemOpen
  \bibfield  {author} {\bibinfo {author} {\bibfnamefont {MP}~\bibnamefont
  {Reijnders}}, \bibinfo {author} {\bibfnamefont {N}~\bibnamefont
  {Debernardi}}, \bibinfo {author} {\bibfnamefont {SB}~\bibnamefont {van~der
  Geer}}, \bibinfo {author} {\bibfnamefont {PHA}\ \bibnamefont {Mutsaers}},
  \bibinfo {author} {\bibfnamefont {EJD}\ \bibnamefont {Vredenbregt}}, and\
  \bibinfo {author} {\bibfnamefont {OJ}~\bibnamefont {Luiten}},\ }\bibfield
  {title} {\bibinfo {title} {Phase-space manipulation of ultracold ion bunches
  with time-dependent fields},\ }\href@noop {} {\bibfield  {journal} {\bibinfo
  {journal} {Physical review letters}\ }\textbf {\bibinfo {volume} {105}},\
  \bibinfo {pages} {034802} (\bibinfo {year} {2010})}\BibitemShut {NoStop}%
\bibitem [{\citenamefont {Oldfield}(1974)}]{oldfield1974microwave}%
  \BibitemOpen
  \bibfield  {author} {\bibinfo {author} {\bibfnamefont {Laurence~Colin}\
  \bibnamefont {Oldfield}},\ }\emph {\bibinfo {title} {Microwave cavities as
  electron lenses.}},\ \href@noop {} {Ph.D. thesis},\ \bibinfo  {school}
  {University of Cambridge} (\bibinfo {year} {1974})\BibitemShut {NoStop}%
\bibitem [{\citenamefont {Chatelain}\ \emph {et~al.}(2012)\citenamefont
  {Chatelain}, \citenamefont {Morrison}, \citenamefont {Godbout},\ and\
  \citenamefont {Siwick}}]{Chatelain2012a}%
  \BibitemOpen
  \bibfield  {author} {\bibinfo {author} {\bibfnamefont {Robert~P}\
  \bibnamefont {Chatelain}}, \bibinfo {author} {\bibfnamefont {Vance~R}\
  \bibnamefont {Morrison}}, \bibinfo {author} {\bibfnamefont {Chris}\
  \bibnamefont {Godbout}}, and\ \bibinfo {author} {\bibfnamefont {Bradley~J}\
  \bibnamefont {Siwick}},\ }\bibfield  {title} {\bibinfo {title} {Ultrafast
  electron diffraction with radio-frequency compressed electron pulses},\
  }\href@noop {} {\bibfield  {journal} {\bibinfo  {journal} {Applied Physics
  Letters}\ }\textbf {\bibinfo {volume} {101}},\ \bibinfo {pages} {081901}
  (\bibinfo {year} {2012})}\BibitemShut {NoStop}%
\bibitem [{\citenamefont {Gliserin}\ \emph {et~al.}(2015)\citenamefont
  {Gliserin}, \citenamefont {Walbran}, \citenamefont {Krausz},\ and\
  \citenamefont {Baum}}]{Gliserin2015}%
  \BibitemOpen
  \bibfield  {author} {\bibinfo {author} {\bibfnamefont {Alexander}\
  \bibnamefont {Gliserin}}, \bibinfo {author} {\bibfnamefont {Matthew}\
  \bibnamefont {Walbran}}, \bibinfo {author} {\bibfnamefont {Ferenc}\
  \bibnamefont {Krausz}}, and\ \bibinfo {author} {\bibfnamefont {Peter}\
  \bibnamefont {Baum}},\ }\bibfield  {title} {\bibinfo {title}
  {Sub-phonon-period compression of electron pulses for atomic diffraction},\
  }\href@noop {} {\bibfield  {journal} {\bibinfo  {journal} {Nature
  communications}\ }\textbf {\bibinfo {volume} {6}},\ \bibinfo {pages} {8723}
  (\bibinfo {year} {2015})}\BibitemShut {NoStop}%
\bibitem [{\citenamefont {Van~Oudheusden}\ \emph {et~al.}(2010)\citenamefont
  {Van~Oudheusden}, \citenamefont {Pasmans}, \citenamefont {Van Der~Geer},
  \citenamefont {De~Loos}, \citenamefont {Van Der~Wiel},\ and\ \citenamefont
  {Luiten}}]{VanOudheusden2010}%
  \BibitemOpen
  \bibfield  {author} {\bibinfo {author} {\bibfnamefont {T}~\bibnamefont
  {Van~Oudheusden}}, \bibinfo {author} {\bibfnamefont {PLEM}\ \bibnamefont
  {Pasmans}}, \bibinfo {author} {\bibfnamefont {SB}~\bibnamefont {Van
  Der~Geer}}, \bibinfo {author} {\bibfnamefont {MJ}~\bibnamefont {De~Loos}},
  \bibinfo {author} {\bibfnamefont {MJ}~\bibnamefont {Van Der~Wiel}}, and\
  \bibinfo {author} {\bibfnamefont {OJ}~\bibnamefont {Luiten}},\ }\bibfield
  {title} {\bibinfo {title} {Compression of subrelativistic
  space-charge-dominated electron bunches for single-shot femtosecond electron
  diffraction},\ }\href@noop {} {\bibfield  {journal} {\bibinfo  {journal}
  {Physical review letters}\ }\textbf {\bibinfo {volume} {105}},\ \bibinfo
  {pages} {264801} (\bibinfo {year} {2010})}\BibitemShut {NoStop}%
\bibitem [{\citenamefont {Maxson}\ \emph {et~al.}(2017)\citenamefont {Maxson},
  \citenamefont {Cesar}, \citenamefont {Calmasini}, \citenamefont {Ody},
  \citenamefont {Musumeci},\ and\ \citenamefont {Alesini}}]{Maxson2017}%
  \BibitemOpen
  \bibfield  {author} {\bibinfo {author} {\bibfnamefont {Jared}\ \bibnamefont
  {Maxson}}, \bibinfo {author} {\bibfnamefont {David}\ \bibnamefont {Cesar}},
  \bibinfo {author} {\bibfnamefont {Giacomo}\ \bibnamefont {Calmasini}},
  \bibinfo {author} {\bibfnamefont {Alexander}\ \bibnamefont {Ody}}, \bibinfo
  {author} {\bibfnamefont {Pietro}\ \bibnamefont {Musumeci}}, and\ \bibinfo
  {author} {\bibfnamefont {David}\ \bibnamefont {Alesini}},\ }\bibfield
  {title} {\bibinfo {title} {Direct measurement of sub-10 fs relativistic
  electron beams with ultralow emittance},\ }\href@noop {} {\bibfield
  {journal} {\bibinfo  {journal} {Physical review letters}\ }\textbf {\bibinfo
  {volume} {118}},\ \bibinfo {pages} {154802} (\bibinfo {year}
  {2017})}\BibitemShut {NoStop}%
\bibitem [{\citenamefont {Zeitler}\ \emph {et~al.}(2015)\citenamefont
  {Zeitler}, \citenamefont {Floettmann},\ and\ \citenamefont
  {Gr{\"u}ner}}]{zeitler2015linearization}%
  \BibitemOpen
  \bibfield  {author} {\bibinfo {author} {\bibfnamefont {Benno}\ \bibnamefont
  {Zeitler}}, \bibinfo {author} {\bibfnamefont {Klaus}\ \bibnamefont
  {Floettmann}}, and\ \bibinfo {author} {\bibfnamefont {Florian}\ \bibnamefont
  {Gr{\"u}ner}},\ }\bibfield  {title} {\bibinfo {title} {Linearization of the
  longitudinal phase space without higher harmonic field},\ }\href@noop {}
  {\bibfield  {journal} {\bibinfo  {journal} {Physical Review Special
  Topics-Accelerators and Beams}\ }\textbf {\bibinfo {volume} {18}},\ \bibinfo
  {pages} {120102} (\bibinfo {year} {2015})}\BibitemShut {NoStop}%
\bibitem [{\citenamefont {Pasmans}\ \emph {et~al.}(2013)\citenamefont
  {Pasmans}, \citenamefont {van~den Ham}, \citenamefont {Dal~Conte},
  \citenamefont {van~der Geer},\ and\ \citenamefont
  {Luiten}}]{pasmans2013microwave}%
  \BibitemOpen
  \bibfield  {author} {\bibinfo {author} {\bibfnamefont {PLEM}\ \bibnamefont
  {Pasmans}}, \bibinfo {author} {\bibfnamefont {GB}~\bibnamefont {van~den
  Ham}}, \bibinfo {author} {\bibfnamefont {SFP}\ \bibnamefont {Dal~Conte}},
  \bibinfo {author} {\bibfnamefont {SB}~\bibnamefont {van~der Geer}}, and\
  \bibinfo {author} {\bibfnamefont {OJ}~\bibnamefont {Luiten}},\ }\bibfield
  {title} {\bibinfo {title} {Microwave tm010 cavities as versatile 4d electron
  optical elements},\ }\href@noop {} {\bibfield  {journal} {\bibinfo  {journal}
  {Ultramicroscopy}\ }\textbf {\bibinfo {volume} {127}},\ \bibinfo {pages}
  {19--24} (\bibinfo {year} {2013})}\BibitemShut {NoStop}%
\bibitem [{\citenamefont {Zhao}\ \emph {et~al.}(2018)\citenamefont {Zhao},
  \citenamefont {Wang}, \citenamefont {Lu}, \citenamefont {Wang}, \citenamefont
  {Hu}, \citenamefont {Wang}, \citenamefont {Qi}, \citenamefont {Jiang},
  \citenamefont {Liu}, \citenamefont {Ma} \emph {et~al.}}]{Zhao2018}%
  \BibitemOpen
  \bibfield  {author} {\bibinfo {author} {\bibfnamefont {Lingrong}\
  \bibnamefont {Zhao}}, \bibinfo {author} {\bibfnamefont {Zhe}\ \bibnamefont
  {Wang}}, \bibinfo {author} {\bibfnamefont {Chao}\ \bibnamefont {Lu}},
  \bibinfo {author} {\bibfnamefont {Rui}\ \bibnamefont {Wang}}, \bibinfo
  {author} {\bibfnamefont {Cheng}\ \bibnamefont {Hu}}, \bibinfo {author}
  {\bibfnamefont {Peng}\ \bibnamefont {Wang}}, \bibinfo {author} {\bibfnamefont
  {Jia}\ \bibnamefont {Qi}}, \bibinfo {author} {\bibfnamefont {Tao}\
  \bibnamefont {Jiang}}, \bibinfo {author} {\bibfnamefont {Shengguang}\
  \bibnamefont {Liu}}, \bibinfo {author} {\bibfnamefont {Zhuoran}\ \bibnamefont
  {Ma}},  \emph {et~al.},\ }\bibfield  {title} {\bibinfo {title} {Terahertz
  streaking of few-femtosecond relativistic electron beams},\ }\href@noop {}
  {\bibfield  {journal} {\bibinfo  {journal} {Physical Review X}\ }\textbf
  {\bibinfo {volume} {8}},\ \bibinfo {pages} {021061} (\bibinfo {year}
  {2018})}\BibitemShut {NoStop}%
\bibitem [{\citenamefont {Li}\ \emph {et~al.}(2019)\citenamefont {Li},
  \citenamefont {Hoffmann}, \citenamefont {Nanni}, \citenamefont {Glenzer},
  \citenamefont {Kozina}, \citenamefont {Lindenberg}, \citenamefont
  {Ofori-Okai}, \citenamefont {Reid}, \citenamefont {Shen}, \citenamefont
  {Weathersby} \emph {et~al.}}]{Li2019}%
  \BibitemOpen
  \bibfield  {author} {\bibinfo {author} {\bibfnamefont {RK}~\bibnamefont
  {Li}}, \bibinfo {author} {\bibfnamefont {MC}~\bibnamefont {Hoffmann}},
  \bibinfo {author} {\bibfnamefont {EA}~\bibnamefont {Nanni}}, \bibinfo
  {author} {\bibfnamefont {SH}~\bibnamefont {Glenzer}}, \bibinfo {author}
  {\bibfnamefont {ME}~\bibnamefont {Kozina}}, \bibinfo {author} {\bibfnamefont
  {AM}~\bibnamefont {Lindenberg}}, \bibinfo {author} {\bibfnamefont
  {BK}~\bibnamefont {Ofori-Okai}}, \bibinfo {author} {\bibfnamefont
  {AH}~\bibnamefont {Reid}}, \bibinfo {author} {\bibfnamefont {X}~\bibnamefont
  {Shen}}, \bibinfo {author} {\bibfnamefont {SP}~\bibnamefont {Weathersby}},
  \emph {et~al.},\ }\bibfield  {title} {\bibinfo {title} {Terahertz-based
  subfemtosecond metrology of relativistic electron beams},\ }\href@noop {}
  {\bibfield  {journal} {\bibinfo  {journal} {Physical Review Accelerators and
  Beams}\ }\textbf {\bibinfo {volume} {22}},\ \bibinfo {pages} {012803}
  (\bibinfo {year} {2019})}\BibitemShut {NoStop}%
\bibitem [{\citenamefont {Cesar}\ and\ \citenamefont
  {Musumeci}(2019)}]{Cesar2019}%
  \BibitemOpen
  \bibfield  {author} {\bibinfo {author} {\bibfnamefont {David}\ \bibnamefont
  {Cesar}}and\ \bibinfo {author} {\bibfnamefont {Pietro}\ \bibnamefont
  {Musumeci}},\ }\bibfield  {title} {\bibinfo {title} {Temporal magnification
  for streaked ultrafast electron diffraction and microscopy},\ }\href@noop {}
  {\bibfield  {journal} {\bibinfo  {journal} {Ultramicroscopy}\ }\textbf
  {\bibinfo {volume} {199}},\ \bibinfo {pages} {1--6} (\bibinfo {year}
  {2019})}\BibitemShut {NoStop}%
\bibitem [{\citenamefont {Kolner}\ and\ \citenamefont
  {Nazarathy}(1990)}]{Kolner1990}%
  \BibitemOpen
  \bibfield  {author} {\bibinfo {author} {\bibfnamefont {B~H}\ \bibnamefont
  {Kolner}}and\ \bibinfo {author} {\bibfnamefont {M}~\bibnamefont
  {Nazarathy}},\ }\bibfield  {title} {\bibinfo {title} {Temporal imaging with a
  time lens.},\ }\href {https://doi.org/10.1364/ol.15.000655} {\bibfield
  {journal} {\bibinfo  {journal} {Optics letters}\ }\textbf {\bibinfo {volume}
  {15}},\ \bibinfo {pages} {655} (\bibinfo {year} {1990})}\BibitemShut
  {NoStop}%
\bibitem [{\citenamefont {Weathersby}\ \emph {et~al.}(2015)\citenamefont
  {Weathersby}, \citenamefont {Brown}, \citenamefont {Centurion}, \citenamefont
  {Chase}, \citenamefont {Coffee}, \citenamefont {Corbett}, \citenamefont
  {Eichner}, \citenamefont {Frisch}, \citenamefont {Fry}, \citenamefont
  {Gühr}, \citenamefont {Hartmann}, \citenamefont {Hast}, \citenamefont
  {Hettel}, \citenamefont {Jobe}, \citenamefont {Jongewaard}, \citenamefont
  {Lewandowski}, \citenamefont {Li}, \citenamefont {Lindenberg}, \citenamefont
  {Makasyuk}, \citenamefont {May}, \citenamefont {McCormick}, \citenamefont
  {Nguyen}, \citenamefont {Reid}, \citenamefont {Shen}, \citenamefont
  {Sokolowski-Tinten}, \citenamefont {Vecchione}, \citenamefont {Vetter},
  \citenamefont {Wu}, \citenamefont {Yang}, \citenamefont {Dürr},\ and\
  \citenamefont {Wang}}]{Weathersby2015}%
  \BibitemOpen
  \bibfield  {author} {\bibinfo {author} {\bibfnamefont {S~P}\ \bibnamefont
  {Weathersby}}, \bibinfo {author} {\bibfnamefont {G}~\bibnamefont {Brown}},
  \bibinfo {author} {\bibfnamefont {M}~\bibnamefont {Centurion}}, \bibinfo
  {author} {\bibfnamefont {T~F}\ \bibnamefont {Chase}}, \bibinfo {author}
  {\bibfnamefont {R}~\bibnamefont {Coffee}}, \bibinfo {author} {\bibfnamefont
  {J}~\bibnamefont {Corbett}}, \bibinfo {author} {\bibfnamefont {J~P}\
  \bibnamefont {Eichner}}, \bibinfo {author} {\bibfnamefont {J~C}\ \bibnamefont
  {Frisch}}, \bibinfo {author} {\bibfnamefont {A~R}\ \bibnamefont {Fry}},
  \bibinfo {author} {\bibfnamefont {M}~\bibnamefont {Gühr}}, \bibinfo {author}
  {\bibfnamefont {N}~\bibnamefont {Hartmann}}, \bibinfo {author} {\bibfnamefont
  {C}~\bibnamefont {Hast}}, \bibinfo {author} {\bibfnamefont {R}~\bibnamefont
  {Hettel}}, \bibinfo {author} {\bibfnamefont {R~K}\ \bibnamefont {Jobe}},
  \bibinfo {author} {\bibfnamefont {E~N}\ \bibnamefont {Jongewaard}}, \bibinfo
  {author} {\bibfnamefont {J~R}\ \bibnamefont {Lewandowski}}, \bibinfo {author}
  {\bibfnamefont {R~K}\ \bibnamefont {Li}}, \bibinfo {author} {\bibfnamefont
  {A~M}\ \bibnamefont {Lindenberg}}, \bibinfo {author} {\bibfnamefont
  {I}~\bibnamefont {Makasyuk}}, \bibinfo {author} {\bibfnamefont {J~E}\
  \bibnamefont {May}}, \bibinfo {author} {\bibfnamefont {D}~\bibnamefont
  {McCormick}}, \bibinfo {author} {\bibfnamefont {M~N}\ \bibnamefont {Nguyen}},
  \bibinfo {author} {\bibfnamefont {A~H}\ \bibnamefont {Reid}}, \bibinfo
  {author} {\bibfnamefont {X}~\bibnamefont {Shen}}, \bibinfo {author}
  {\bibfnamefont {K}~\bibnamefont {Sokolowski-Tinten}}, \bibinfo {author}
  {\bibfnamefont {T}~\bibnamefont {Vecchione}}, \bibinfo {author}
  {\bibfnamefont {S~L}\ \bibnamefont {Vetter}}, \bibinfo {author}
  {\bibfnamefont {J}~\bibnamefont {Wu}}, \bibinfo {author} {\bibfnamefont
  {J}~\bibnamefont {Yang}}, \bibinfo {author} {\bibfnamefont {H~A}\
  \bibnamefont {Dürr}}, and\ \bibinfo {author} {\bibfnamefont {X~J}\
  \bibnamefont {Wang}},\ }\bibfield  {title} {\bibinfo {title}
  {Mega-electron-volt ultrafast electron diffraction at slac national
  accelerator laboratory.},\ }\href {https://doi.org/10.1063/1.4926994}
  {\bibfield  {journal} {\bibinfo  {journal} {The Review of scientific
  instruments}\ }\textbf {\bibinfo {volume} {86}},\ \bibinfo {pages} {073702}
  (\bibinfo {year} {2015})}\BibitemShut {NoStop}%
\bibitem [{\citenamefont {Hachmann}\ and\ \citenamefont
  {Flöttmann}(2016)}]{Hachmann2016}%
  \BibitemOpen
  \bibfield  {author} {\bibinfo {author} {\bibfnamefont {M.}~\bibnamefont
  {Hachmann}}and\ \bibinfo {author} {\bibfnamefont {K.}~\bibnamefont
  {Flöttmann}},\ }\bibfield  {title} {\bibinfo {title} {{Measurement of ultra
  low transverse emittance at REGAE}},\ }\bibfield  {booktitle} {\emph
  {\bibinfo {booktitle} {{Proceedings, 2nd European Advanced Accelerator
  Concepts Workshop (EAAC2015): La Biodola, Isola d'Elba, Italy, September
  12-19, 2015}}},\ }\href {https://doi.org/10.1016/j.nima.2016.01.065}
  {\bibfield  {journal} {\bibinfo  {journal} {Nucl. Instrum. Meth.}\ }\textbf
  {\bibinfo {volume} {A829}},\ \bibinfo {pages} {318--320} (\bibinfo {year}
  {2016})}\BibitemShut {NoStop}%
%%CITATION = NUIMA,A829,318;%%
\bibitem [{\citenamefont {Filippetto}\ and\ \citenamefont
  {Qian}(2016)}]{Filippetto2016}%
  \BibitemOpen
  \bibfield  {author} {\bibinfo {author} {\bibfnamefont {D}~\bibnamefont
  {Filippetto}}and\ \bibinfo {author} {\bibfnamefont {H}~\bibnamefont {Qian}},\
  }\bibfield  {title} {\bibinfo {title} {Design of a high-flux instrument for
  ultrafast electron diffraction and microscopy},\ }\href@noop {} {\bibfield
  {journal} {\bibinfo  {journal} {Journal of Physics B: Atomic, Molecular and
  Optical Physics}\ }\textbf {\bibinfo {volume} {49}},\ \bibinfo {pages}
  {104003} (\bibinfo {year} {2016})}\BibitemShut {NoStop}%
\bibitem [{\citenamefont {Sannomiya}\ \emph {et~al.}(2019)\citenamefont
  {Sannomiya}, \citenamefont {Arai}, \citenamefont {Nagayama},\ and\
  \citenamefont {Nagatani}}]{Sannomiya2019}%
  \BibitemOpen
  \bibfield  {author} {\bibinfo {author} {\bibfnamefont {Takumi}\ \bibnamefont
  {Sannomiya}}, \bibinfo {author} {\bibfnamefont {Yoshihiro}\ \bibnamefont
  {Arai}}, \bibinfo {author} {\bibfnamefont {Kuniaki}\ \bibnamefont
  {Nagayama}}, and\ \bibinfo {author} {\bibfnamefont {Yukinori}\ \bibnamefont
  {Nagatani}},\ }\bibfield  {title} {\bibinfo {title} {Transmission electron
  microscope using a linear accelerator},\ }\href@noop {} {\bibfield  {journal}
  {\bibinfo  {journal} {Physical review letters}\ }\textbf {\bibinfo {volume}
  {123}},\ \bibinfo {pages} {150801} (\bibinfo {year} {2019})}\BibitemShut
  {NoStop}%
\bibitem [{\citenamefont {Lassise}\ \emph {et~al.}(2012)\citenamefont
  {Lassise}, \citenamefont {Mutsaers},\ and\ \citenamefont
  {Luiten}}]{Lassise2012}%
  \BibitemOpen
  \bibfield  {author} {\bibinfo {author} {\bibfnamefont {A}~\bibnamefont
  {Lassise}}, \bibinfo {author} {\bibfnamefont {PHA}\ \bibnamefont {Mutsaers}},
  and\ \bibinfo {author} {\bibfnamefont {OJ}~\bibnamefont {Luiten}},\
  }\bibfield  {title} {\bibinfo {title} {Compact, low power radio frequency
  cavity for femtosecond electron microscopy},\ }\href@noop {} {\bibfield
  {journal} {\bibinfo  {journal} {Review of scientific instruments}\ }\textbf
  {\bibinfo {volume} {83}},\ \bibinfo {pages} {043705} (\bibinfo {year}
  {2012})}\BibitemShut {NoStop}%
\bibitem [{\citenamefont {Oldfield}(1976)}]{Oldfield1976}%
  \BibitemOpen
  \bibfield  {author} {\bibinfo {author} {\bibfnamefont {Laurence~C}\
  \bibnamefont {Oldfield}},\ }\bibfield  {title} {\bibinfo {title} {A
  rotationally symmetric electron beam chopper for picosecond pulses},\
  }\href@noop {} {\bibfield  {journal} {\bibinfo  {journal} {Journal of Physics
  E: Scientific Instruments}\ }\textbf {\bibinfo {volume} {9}},\ \bibinfo
  {pages} {455} (\bibinfo {year} {1976})}\BibitemShut {NoStop}%
\bibitem [{\citenamefont {Verhoeven}\ \emph
  {et~al.}(2018{\natexlab{a}})\citenamefont {Verhoeven}, \citenamefont {van
  Rens}, \citenamefont {Kieft}, \citenamefont {Mutsaers},\ and\ \citenamefont
  {Luiten}}]{Verhoeven2018}%
  \BibitemOpen
  \bibfield  {author} {\bibinfo {author} {\bibfnamefont {W}~\bibnamefont
  {Verhoeven}}, \bibinfo {author} {\bibfnamefont {J~F~M}\ \bibnamefont {van
  Rens}}, \bibinfo {author} {\bibfnamefont {E~R}\ \bibnamefont {Kieft}},
  \bibinfo {author} {\bibfnamefont {P~H~A}\ \bibnamefont {Mutsaers}}, and\
  \bibinfo {author} {\bibfnamefont {O~J}\ \bibnamefont {Luiten}},\ }\bibfield
  {title} {\bibinfo {title} {High quality ultrafast transmission electron
  microscopy using resonant microwave cavities.},\ }\href
  {https://doi.org/10.1016/j.ultramic.2018.03.012} {\bibfield  {journal}
  {\bibinfo  {journal} {Ultramicroscopy}\ }\textbf {\bibinfo {volume} {188}},\
  \bibinfo {pages} {85--89} (\bibinfo {year} {2018}{\natexlab{a}})}\BibitemShut
  {NoStop}%
\bibitem [{\citenamefont {Murooka}\ \emph {et~al.}(2011)\citenamefont
  {Murooka}, \citenamefont {Naruse}, \citenamefont {Sakakihara}, \citenamefont
  {Ishimaru}, \citenamefont {Yang},\ and\ \citenamefont
  {Tanimura}}]{murooka2011transmission}%
  \BibitemOpen
  \bibfield  {author} {\bibinfo {author} {\bibfnamefont {Y}~\bibnamefont
  {Murooka}}, \bibinfo {author} {\bibfnamefont {N}~\bibnamefont {Naruse}},
  \bibinfo {author} {\bibfnamefont {S}~\bibnamefont {Sakakihara}}, \bibinfo
  {author} {\bibfnamefont {M}~\bibnamefont {Ishimaru}}, \bibinfo {author}
  {\bibfnamefont {J}~\bibnamefont {Yang}}, and\ \bibinfo {author}
  {\bibfnamefont {K}~\bibnamefont {Tanimura}},\ }\bibfield  {title} {\bibinfo
  {title} {Transmission-electron diffraction by mev electron pulses},\
  }\href@noop {} {\bibfield  {journal} {\bibinfo  {journal} {Applied Physics
  Letters}\ }\textbf {\bibinfo {volume} {98}},\ \bibinfo {pages} {251903}
  (\bibinfo {year} {2011})}\BibitemShut {NoStop}%
\bibitem [{\citenamefont {Verhoeven}\ \emph
  {et~al.}(2018{\natexlab{b}})\citenamefont {Verhoeven}, \citenamefont {van
  Rens}, \citenamefont {Toonen}, \citenamefont {Kieft}, \citenamefont
  {Mutsaers},\ and\ \citenamefont {Luiten}}]{Verhoeven}%
  \BibitemOpen
  \bibfield  {author} {\bibinfo {author} {\bibfnamefont {W.}~\bibnamefont
  {Verhoeven}}, \bibinfo {author} {\bibfnamefont {J.~F.~M.}\ \bibnamefont {van
  Rens}}, \bibinfo {author} {\bibfnamefont {W.~F.}\ \bibnamefont {Toonen}},
  \bibinfo {author} {\bibfnamefont {E.~R.}\ \bibnamefont {Kieft}}, \bibinfo
  {author} {\bibfnamefont {P.~H.~A.}\ \bibnamefont {Mutsaers}}, and\ \bibinfo
  {author} {\bibfnamefont {O.~J.}\ \bibnamefont {Luiten}},\ }\bibfield  {title}
  {\bibinfo {title} {Time-of-flight electron energy loss spectroscopy by
  longitudinal phase space manipulation with microwave cavities},\ }\href
  {https://doi.org/10.1063/1.5052217} {\bibfield  {journal} {\bibinfo
  {journal} {Structural Dynamics}\ }\textbf {\bibinfo {volume} {5}},\ \bibinfo
  {pages} {051101} (\bibinfo {year} {2018}{\natexlab{b}})}\BibitemShut
  {NoStop}%
\bibitem [{\citenamefont {Ura}\ and\ \citenamefont
  {Morimura}(1973)}]{ura1973generation}%
  \BibitemOpen
  \bibfield  {author} {\bibinfo {author} {\bibfnamefont {K}~\bibnamefont
  {Ura}}and\ \bibinfo {author} {\bibfnamefont {N}~\bibnamefont {Morimura}},\
  }\bibfield  {title} {\bibinfo {title} {Generation of picosecond pulse
  electron beams},\ }\href@noop {} {\bibfield  {journal} {\bibinfo  {journal}
  {Journal of Vacuum Science and Technology}\ }\textbf {\bibinfo {volume}
  {10}},\ \bibinfo {pages} {948--950} (\bibinfo {year} {1973})}\BibitemShut
  {NoStop}%
\bibitem [{\citenamefont {Hosokawa}\ \emph {et~al.}(1978)\citenamefont
  {Hosokawa}, \citenamefont {Fujioka},\ and\ \citenamefont
  {Ura}}]{hosokawa1978gigahertz}%
  \BibitemOpen
  \bibfield  {author} {\bibinfo {author} {\bibfnamefont {T}~\bibnamefont
  {Hosokawa}}, \bibinfo {author} {\bibfnamefont {H}~\bibnamefont {Fujioka}},
  and\ \bibinfo {author} {\bibfnamefont {K}~\bibnamefont {Ura}},\ }\bibfield
  {title} {\bibinfo {title} {Gigahertz stroboscopy with the scanning electron
  microscope},\ }\href@noop {} {\bibfield  {journal} {\bibinfo  {journal}
  {Review of Scientific Instruments}\ }\textbf {\bibinfo {volume} {49}},\
  \bibinfo {pages} {1293--1299} (\bibinfo {year} {1978})}\BibitemShut {NoStop}%
\bibitem [{\citenamefont {Otto}\ \emph {et~al.}(2017)\citenamefont {Otto},
  \citenamefont {Ren{\'e}~de Cotret}, \citenamefont {Stern},\ and\
  \citenamefont {Siwick}}]{otto2017solving}%
  \BibitemOpen
  \bibfield  {author} {\bibinfo {author} {\bibfnamefont {Martin~R}\
  \bibnamefont {Otto}}, \bibinfo {author} {\bibfnamefont {LP}~\bibnamefont
  {Ren{\'e}~de Cotret}}, \bibinfo {author} {\bibfnamefont {Mark~J}\
  \bibnamefont {Stern}}, and\ \bibinfo {author} {\bibfnamefont {Bradley~J}\
  \bibnamefont {Siwick}},\ }\bibfield  {title} {\bibinfo {title} {Solving the
  jitter problem in microwave compressed ultrafast electron diffraction
  instruments: Robust sub-50 fs cavity-laser phase stabilization},\ }\href@noop
  {} {\bibfield  {journal} {\bibinfo  {journal} {Structural Dynamics}\ }\textbf
  {\bibinfo {volume} {4}},\ \bibinfo {pages} {051101} (\bibinfo {year}
  {2017})}\BibitemShut {NoStop}%
\bibitem [{\citenamefont {Krivanek}\ \emph {et~al.}(2014)\citenamefont
  {Krivanek}, \citenamefont {Lovejoy}, \citenamefont {Dellby}, \citenamefont
  {Aoki}, \citenamefont {Carpenter}, \citenamefont {Rez}, \citenamefont
  {Soignard}, \citenamefont {Zhu}, \citenamefont {Batson}, \citenamefont
  {Lagos} \emph {et~al.}}]{Krivanek2014}%
  \BibitemOpen
  \bibfield  {author} {\bibinfo {author} {\bibfnamefont {Ondrej~L}\
  \bibnamefont {Krivanek}}, \bibinfo {author} {\bibfnamefont {Tracy~C}\
  \bibnamefont {Lovejoy}}, \bibinfo {author} {\bibfnamefont {Niklas}\
  \bibnamefont {Dellby}}, \bibinfo {author} {\bibfnamefont {Toshihiro}\
  \bibnamefont {Aoki}}, \bibinfo {author} {\bibfnamefont {RW}~\bibnamefont
  {Carpenter}}, \bibinfo {author} {\bibfnamefont {Peter}\ \bibnamefont {Rez}},
  \bibinfo {author} {\bibfnamefont {Emmanuel}\ \bibnamefont {Soignard}},
  \bibinfo {author} {\bibfnamefont {Jiangtao}\ \bibnamefont {Zhu}}, \bibinfo
  {author} {\bibfnamefont {Philip~E}\ \bibnamefont {Batson}}, \bibinfo {author}
  {\bibfnamefont {Maureen~J}\ \bibnamefont {Lagos}},  \emph {et~al.},\
  }\bibfield  {title} {\bibinfo {title} {Vibrational spectroscopy in the
  electron microscope},\ }\href@noop {} {\bibfield  {journal} {\bibinfo
  {journal} {Nature}\ }\textbf {\bibinfo {volume} {514}},\ \bibinfo {pages}
  {209} (\bibinfo {year} {2014})}\BibitemShut {NoStop}%
\bibitem [{\citenamefont {Venkatraman}\ \emph {et~al.}(2019)\citenamefont
  {Venkatraman}, \citenamefont {Levin}, \citenamefont {March}, \citenamefont
  {Rez},\ and\ \citenamefont {Crozier}}]{venkatraman2019vibrational}%
  \BibitemOpen
  \bibfield  {author} {\bibinfo {author} {\bibfnamefont {Kartik}\ \bibnamefont
  {Venkatraman}}, \bibinfo {author} {\bibfnamefont {Barnaby~DA}\ \bibnamefont
  {Levin}}, \bibinfo {author} {\bibfnamefont {Katia}\ \bibnamefont {March}},
  \bibinfo {author} {\bibfnamefont {Peter}\ \bibnamefont {Rez}}, and\ \bibinfo
  {author} {\bibfnamefont {Peter~A}\ \bibnamefont {Crozier}},\ }\bibfield
  {title} {\bibinfo {title} {Vibrational spectroscopy at atomic resolution with
  electron impact scattering},\ }\href@noop {} {\bibfield  {journal} {\bibinfo
  {journal} {Nature Physics}\ }\textbf {\bibinfo {volume} {15}},\ \bibinfo
  {pages} {1237--1241} (\bibinfo {year} {2019})}\BibitemShut {NoStop}%
\bibitem [{\citenamefont {Pomarico}\ \emph {et~al.}(2018)\citenamefont
  {Pomarico}, \citenamefont {Kim}, \citenamefont {García~de Abajo},
  \citenamefont {Kwon}, \citenamefont {Carbone},\ and\ \citenamefont {van~der
  Veen}}]{carbonereview}%
  \BibitemOpen
  \bibfield  {author} {\bibinfo {author} {\bibfnamefont {Enrico}\ \bibnamefont
  {Pomarico}}, \bibinfo {author} {\bibfnamefont {Ye-Jin}\ \bibnamefont {Kim}},
  \bibinfo {author} {\bibfnamefont {F.~Javier}\ \bibnamefont {García~de
  Abajo}}, \bibinfo {author} {\bibfnamefont {Oh-Hoon}\ \bibnamefont {Kwon}},
  \bibinfo {author} {\bibfnamefont {Fabrizio}\ \bibnamefont {Carbone}}, and\
  \bibinfo {author} {\bibfnamefont {Renske~M.}\ \bibnamefont {van~der Veen}},\
  }\bibfield  {title} {\bibinfo {title} {Ultrafast electron energy-loss
  spectroscopy in transmission electron microscopy},\ }\href
  {https://doi.org/10.1557/mrs.2018.148} {\bibfield  {journal} {\bibinfo
  {journal} {MRS Bulletin}\ }\textbf {\bibinfo {volume} {43}},\ \bibinfo
  {pages} {497–503} (\bibinfo {year} {2018})}\BibitemShut {NoStop}%
\bibitem [{\citenamefont {Joy}\ and\ \citenamefont {Joy}(1996)}]{joy1996low}%
  \BibitemOpen
  \bibfield  {author} {\bibinfo {author} {\bibfnamefont {David~C}\ \bibnamefont
  {Joy}}and\ \bibinfo {author} {\bibfnamefont {Carolyn~S}\ \bibnamefont
  {Joy}},\ }\bibfield  {title} {\bibinfo {title} {Low voltage scanning electron
  microscopy},\ }\href@noop {} {\bibfield  {journal} {\bibinfo  {journal}
  {Micron}\ }\textbf {\bibinfo {volume} {27}},\ \bibinfo {pages} {247--263}
  (\bibinfo {year} {1996})}\BibitemShut {NoStop}%
\bibitem [{\citenamefont {Koppell}\ \emph {et~al.}(2019)\citenamefont
  {Koppell}, \citenamefont {Mankos}, \citenamefont {Bowman}, \citenamefont
  {Israel}, \citenamefont {Juffmann}, \citenamefont {Klopfer},\ and\
  \citenamefont {Kasevich}}]{Koppell2019}%
  \BibitemOpen
  \bibfield  {author} {\bibinfo {author} {\bibfnamefont {Stewart~A}\
  \bibnamefont {Koppell}}, \bibinfo {author} {\bibfnamefont {Marian}\
  \bibnamefont {Mankos}}, \bibinfo {author} {\bibfnamefont {Adam~J}\
  \bibnamefont {Bowman}}, \bibinfo {author} {\bibfnamefont {Yonatan}\
  \bibnamefont {Israel}}, \bibinfo {author} {\bibfnamefont {Thomas}\
  \bibnamefont {Juffmann}}, \bibinfo {author} {\bibfnamefont {Brannon~B}\
  \bibnamefont {Klopfer}}, and\ \bibinfo {author} {\bibfnamefont {Mark~A}\
  \bibnamefont {Kasevich}},\ }\bibfield  {title} {\bibinfo {title} {Design for
  a 10 kev multi-pass transmission electron microscope},\ }\href@noop {}
  {\bibfield  {journal} {\bibinfo  {journal} {arXiv preprint arXiv:1904.11064}\
  } (\bibinfo {year} {2019})}\BibitemShut {NoStop}%
\bibitem [{\citenamefont {Krivanek}\ \emph {et~al.}(2009)\citenamefont
  {Krivanek}, \citenamefont {Ursin}, \citenamefont {Bacon}, \citenamefont
  {Corbin}, \citenamefont {Dellby}, \citenamefont {Hrncirik}, \citenamefont
  {Murfitt}, \citenamefont {Own},\ and\ \citenamefont
  {Szilagyi}}]{krivanek2009high}%
  \BibitemOpen
  \bibfield  {author} {\bibinfo {author} {\bibfnamefont {Ondrej~L}\
  \bibnamefont {Krivanek}}, \bibinfo {author} {\bibfnamefont {Jonathan~P}\
  \bibnamefont {Ursin}}, \bibinfo {author} {\bibfnamefont {Neil~J}\
  \bibnamefont {Bacon}}, \bibinfo {author} {\bibfnamefont {George~J}\
  \bibnamefont {Corbin}}, \bibinfo {author} {\bibfnamefont {Niklas}\
  \bibnamefont {Dellby}}, \bibinfo {author} {\bibfnamefont {Petr}\ \bibnamefont
  {Hrncirik}}, \bibinfo {author} {\bibfnamefont {Matthew~F}\ \bibnamefont
  {Murfitt}}, \bibinfo {author} {\bibfnamefont {Christopher~S}\ \bibnamefont
  {Own}}, and\ \bibinfo {author} {\bibfnamefont {Zoltan~S}\ \bibnamefont
  {Szilagyi}},\ }\bibfield  {title} {\bibinfo {title} {High-energy-resolution
  monochromator for aberration-corrected scanning transmission electron
  microscopy/electron energy-loss spectroscopy},\ }\href@noop {} {\bibfield
  {journal} {\bibinfo  {journal} {Philosophical Transactions of the Royal
  Society A: Mathematical, Physical and Engineering Sciences}\ }\textbf
  {\bibinfo {volume} {367}},\ \bibinfo {pages} {3683--3697} (\bibinfo {year}
  {2009})}\BibitemShut {NoStop}%
\bibitem [{\citenamefont {Mankos}\ \emph {et~al.}(2016)\citenamefont {Mankos},
  \citenamefont {Shadman},\ and\ \citenamefont {Kolarik}}]{Mankos2016}%
  \BibitemOpen
  \bibfield  {author} {\bibinfo {author} {\bibfnamefont {Marian}\ \bibnamefont
  {Mankos}}, \bibinfo {author} {\bibfnamefont {Khashayar}\ \bibnamefont
  {Shadman}}, and\ \bibinfo {author} {\bibfnamefont {Vladimir}\ \bibnamefont
  {Kolarik}},\ }\bibfield  {title} {\bibinfo {title} {Novel electron
  monochromator for high resolution imaging and spectroscopy},\ }\href@noop {}
  {\bibfield  {journal} {\bibinfo  {journal} {Journal of Vacuum Science \&
  Technology B, Nanotechnology and Microelectronics: Materials, Processing,
  Measurement, and Phenomena}\ }\textbf {\bibinfo {volume} {34}},\ \bibinfo
  {pages} {06KP01} (\bibinfo {year} {2016})}\BibitemShut {NoStop}%
\bibitem [{\citenamefont {Williams}\ \emph {et~al.}(2017)\citenamefont
  {Williams}, \citenamefont {Zhou}, \citenamefont {Sun}, \citenamefont {Tao},
  \citenamefont {Chang}, \citenamefont {Makino}, \citenamefont {Berz},
  \citenamefont {Duxbury},\ and\ \citenamefont {Ruan}}]{Williams2017}%
  \BibitemOpen
  \bibfield  {author} {\bibinfo {author} {\bibfnamefont {J}~\bibnamefont
  {Williams}}, \bibinfo {author} {\bibfnamefont {F}~\bibnamefont {Zhou}},
  \bibinfo {author} {\bibfnamefont {T}~\bibnamefont {Sun}}, \bibinfo {author}
  {\bibfnamefont {Z}~\bibnamefont {Tao}}, \bibinfo {author} {\bibfnamefont
  {K}~\bibnamefont {Chang}}, \bibinfo {author} {\bibfnamefont {K}~\bibnamefont
  {Makino}}, \bibinfo {author} {\bibfnamefont {M}~\bibnamefont {Berz}},
  \bibinfo {author} {\bibfnamefont {PM}~\bibnamefont {Duxbury}}, and\ \bibinfo
  {author} {\bibfnamefont {C-Y}\ \bibnamefont {Ruan}},\ }\bibfield  {title}
  {\bibinfo {title} {Active control of bright electron beams with rf optics for
  femtosecond microscopy},\ }\href@noop {} {\bibfield  {journal} {\bibinfo
  {journal} {Structural Dynamics}\ }\textbf {\bibinfo {volume} {4}},\ \bibinfo
  {pages} {044035} (\bibinfo {year} {2017})}\BibitemShut {NoStop}%
\bibitem [{\citenamefont {Li}\ and\ \citenamefont {Wang}(2017)}]{Li2017}%
  \BibitemOpen
  \bibfield  {author} {\bibinfo {author} {\bibfnamefont {RK}~\bibnamefont
  {Li}}and\ \bibinfo {author} {\bibfnamefont {XJ}~\bibnamefont {Wang}},\
  }\bibfield  {title} {\bibinfo {title} {Femtosecond mev electron energy-loss
  spectroscopy},\ }\href@noop {} {\bibfield  {journal} {\bibinfo  {journal}
  {Physical Review Applied}\ }\textbf {\bibinfo {volume} {8}},\ \bibinfo
  {pages} {054017} (\bibinfo {year} {2017})}\BibitemShut {NoStop}%
\bibitem [{\citenamefont {Karkare}\ \emph {et~al.}(2014)\citenamefont
  {Karkare}, \citenamefont {Boulet}, \citenamefont {Cultrera}, \citenamefont
  {Dunham}, \citenamefont {Liu}, \citenamefont {Schaff},\ and\ \citenamefont
  {Bazarov}}]{Karkare2014}%
  \BibitemOpen
  \bibfield  {author} {\bibinfo {author} {\bibfnamefont {Siddharth}\
  \bibnamefont {Karkare}}, \bibinfo {author} {\bibfnamefont {Laurent}\
  \bibnamefont {Boulet}}, \bibinfo {author} {\bibfnamefont {Luca}\ \bibnamefont
  {Cultrera}}, \bibinfo {author} {\bibfnamefont {Bruce}\ \bibnamefont
  {Dunham}}, \bibinfo {author} {\bibfnamefont {Xianghong}\ \bibnamefont {Liu}},
  \bibinfo {author} {\bibfnamefont {William}\ \bibnamefont {Schaff}}, and\
  \bibinfo {author} {\bibfnamefont {Ivan}\ \bibnamefont {Bazarov}},\ }\bibfield
   {title} {\bibinfo {title} {Ultrabright and ultrafast iii-v semiconductor
  photocathodes.},\ }\href {https://doi.org/10.1103/PhysRevLett.112.097601}
  {\bibfield  {journal} {\bibinfo  {journal} {Physical review letters}\
  }\textbf {\bibinfo {volume} {112}},\ \bibinfo {pages} {097601} (\bibinfo
  {year} {2014})}\BibitemShut {NoStop}%
\bibitem [{\citenamefont {Rao}\ and\ \citenamefont {Dowell}()}]{Rao2014}%
  \BibitemOpen
  \bibfield  {author} {\bibinfo {author} {\bibfnamefont {Triveni}\ \bibnamefont
  {Rao}}and\ \bibinfo {author} {\bibfnamefont {David~H.}\ \bibnamefont
  {Dowell}},\ }\bibfield  {title} {\bibinfo {title} {An engineering guide to
  photoinjectors},\ }\href@noop {} {\ }\Eprint
  {https://arxiv.org/abs/http://arxiv.org/abs/1403.7539v1}
  {http://arxiv.org/abs/1403.7539v1} \BibitemShut {NoStop}%
\bibitem [{\citenamefont {Bartels}\ \emph {et~al.}(1999)\citenamefont
  {Bartels}, \citenamefont {Dekorsy},\ and\ \citenamefont
  {Kurz}}]{Bartels1999}%
  \BibitemOpen
  \bibfield  {author} {\bibinfo {author} {\bibfnamefont {Albrecht}\
  \bibnamefont {Bartels}}, \bibinfo {author} {\bibfnamefont {Thomas}\
  \bibnamefont {Dekorsy}}, and\ \bibinfo {author} {\bibfnamefont {Heinrich}\
  \bibnamefont {Kurz}},\ }\bibfield  {title} {\bibinfo {title} {Femtosecond ti:
  sapphire ring laser with a 2-ghz repetition rate and its application in
  time-resolved spectroscopy},\ }\href@noop {} {\bibfield  {journal} {\bibinfo
  {journal} {Optics letters}\ }\textbf {\bibinfo {volume} {24}},\ \bibinfo
  {pages} {996--998} (\bibinfo {year} {1999})}\BibitemShut {NoStop}%
\bibitem [{\citenamefont {Krivanek}\ \emph {et~al.}(2013)\citenamefont
  {Krivanek}, \citenamefont {Lovejoy}, \citenamefont {Dellby},\ and\
  \citenamefont {Carpenter}}]{krivanek2013monochromated}%
  \BibitemOpen
  \bibfield  {author} {\bibinfo {author} {\bibfnamefont {Ondrej~L}\
  \bibnamefont {Krivanek}}, \bibinfo {author} {\bibfnamefont {Tracy~C}\
  \bibnamefont {Lovejoy}}, \bibinfo {author} {\bibfnamefont {Niklas}\
  \bibnamefont {Dellby}}, and\ \bibinfo {author} {\bibfnamefont
  {RW}~\bibnamefont {Carpenter}},\ }\bibfield  {title} {\bibinfo {title}
  {Monochromated stem with a 30 mev-wide, atom-sized electron probe},\
  }\href@noop {} {\bibfield  {journal} {\bibinfo  {journal} {Microscopy}\
  }\textbf {\bibinfo {volume} {62}},\ \bibinfo {pages} {3--21} (\bibinfo {year}
  {2013})}\BibitemShut {NoStop}%
\bibitem [{\citenamefont {Dowell}\ and\ \citenamefont
  {Schmerge}(2009)}]{Dowell2009}%
  \BibitemOpen
  \bibfield  {author} {\bibinfo {author} {\bibfnamefont {DH}~\bibnamefont
  {Dowell}}and\ \bibinfo {author} {\bibfnamefont {JF}~\bibnamefont
  {Schmerge}},\ }\bibfield  {title} {\bibinfo {title} {Quantum efficiency and
  thermal emittance of metal photocathodes},\ }\href@noop {} {\bibfield
  {journal} {\bibinfo  {journal} {Physical Review Special Topics-Accelerators
  and Beams}\ }\textbf {\bibinfo {volume} {12}},\ \bibinfo {pages} {074201}
  (\bibinfo {year} {2009})}\BibitemShut {NoStop}%
\bibitem [{\citenamefont {Feist}\ \emph {et~al.}(2017)\citenamefont {Feist},
  \citenamefont {Bach}, \citenamefont {da~Silva}, \citenamefont {Danz},
  \citenamefont {M{\"o}ller}, \citenamefont {Priebe}, \citenamefont
  {Domr{\"o}se}, \citenamefont {Gatzmann}, \citenamefont {Rost}, \citenamefont
  {Schauss} \emph {et~al.}}]{feist2017ultrafast}%
  \BibitemOpen
  \bibfield  {author} {\bibinfo {author} {\bibfnamefont {Armin}\ \bibnamefont
  {Feist}}, \bibinfo {author} {\bibfnamefont {Nora}\ \bibnamefont {Bach}},
  \bibinfo {author} {\bibfnamefont {Nara~Rubiano}\ \bibnamefont {da~Silva}},
  \bibinfo {author} {\bibfnamefont {Thomas}\ \bibnamefont {Danz}}, \bibinfo
  {author} {\bibfnamefont {Marcel}\ \bibnamefont {M{\"o}ller}}, \bibinfo
  {author} {\bibfnamefont {Katharina~E}\ \bibnamefont {Priebe}}, \bibinfo
  {author} {\bibfnamefont {Till}\ \bibnamefont {Domr{\"o}se}}, \bibinfo
  {author} {\bibfnamefont {J~Gregor}\ \bibnamefont {Gatzmann}}, \bibinfo
  {author} {\bibfnamefont {Stefan}\ \bibnamefont {Rost}}, \bibinfo {author}
  {\bibfnamefont {Jakob}\ \bibnamefont {Schauss}},  \emph {et~al.},\ }\bibfield
   {title} {\bibinfo {title} {Ultrafast transmission electron microscopy using
  a laser-driven field emitter: Femtosecond resolution with a high coherence
  electron beam},\ }\href@noop {} {\bibfield  {journal} {\bibinfo  {journal}
  {Ultramicroscopy}\ }\textbf {\bibinfo {volume} {176}},\ \bibinfo {pages}
  {63--73} (\bibinfo {year} {2017})}\BibitemShut {NoStop}%
\bibitem [{\citenamefont {Scheinfein}\ \emph {et~al.}(1993)\citenamefont
  {Scheinfein}, \citenamefont {Qian},\ and\ \citenamefont
  {Spence}}]{Scheinfein1993}%
  \BibitemOpen
  \bibfield  {author} {\bibinfo {author} {\bibfnamefont {MR}~\bibnamefont
  {Scheinfein}}, \bibinfo {author} {\bibfnamefont {W}~\bibnamefont {Qian}},
  and\ \bibinfo {author} {\bibfnamefont {JCH}\ \bibnamefont {Spence}},\
  }\bibfield  {title} {\bibinfo {title} {Aberrations of emission cathodes:
  Nanometer diameter field-emission electron sources},\ }\href@noop {}
  {\bibfield  {journal} {\bibinfo  {journal} {Journal of applied physics}\
  }\textbf {\bibinfo {volume} {73}},\ \bibinfo {pages} {2057--2068} (\bibinfo
  {year} {1993})}\BibitemShut {NoStop}%
\bibitem [{\citenamefont {Stern}\ \emph {et~al.}(2018)\citenamefont {Stern},
  \citenamefont {de~Cotret}, \citenamefont {Otto}, \citenamefont {Chatelain},
  \citenamefont {Boisvert}, \citenamefont {Sutton},\ and\ \citenamefont
  {Siwick}}]{Stern2018}%
  \BibitemOpen
  \bibfield  {author} {\bibinfo {author} {\bibfnamefont {Mark~J}\ \bibnamefont
  {Stern}}, \bibinfo {author} {\bibfnamefont {Laurent P~Ren{\'e}}\ \bibnamefont
  {de~Cotret}}, \bibinfo {author} {\bibfnamefont {Martin~R}\ \bibnamefont
  {Otto}}, \bibinfo {author} {\bibfnamefont {Robert~P}\ \bibnamefont
  {Chatelain}}, \bibinfo {author} {\bibfnamefont {Jean-Philippe}\ \bibnamefont
  {Boisvert}}, \bibinfo {author} {\bibfnamefont {Mark}\ \bibnamefont {Sutton}},
  and\ \bibinfo {author} {\bibfnamefont {Bradley~J}\ \bibnamefont {Siwick}},\
  }\bibfield  {title} {\bibinfo {title} {Mapping momentum-dependent
  electron-phonon coupling and nonequilibrium phonon dynamics with ultrafast
  electron diffuse scattering},\ }\href@noop {} {\bibfield  {journal} {\bibinfo
   {journal} {Physical Review B}\ }\textbf {\bibinfo {volume} {97}},\ \bibinfo
  {pages} {165416} (\bibinfo {year} {2018})}\BibitemShut {NoStop}%
\bibitem [{\citenamefont {Cultrera}\ \emph {et~al.}(2015)\citenamefont
  {Cultrera}, \citenamefont {Karkare}, \citenamefont {Lee}, \citenamefont
  {Liu}, \citenamefont {Bazarov},\ and\ \citenamefont
  {Dunham}}]{cultrera2015cold}%
  \BibitemOpen
  \bibfield  {author} {\bibinfo {author} {\bibfnamefont {Luca}\ \bibnamefont
  {Cultrera}}, \bibinfo {author} {\bibfnamefont {Siddharth}\ \bibnamefont
  {Karkare}}, \bibinfo {author} {\bibfnamefont {Hyeri}\ \bibnamefont {Lee}},
  \bibinfo {author} {\bibfnamefont {Xianghong}\ \bibnamefont {Liu}}, \bibinfo
  {author} {\bibfnamefont {Ivan}\ \bibnamefont {Bazarov}}, and\ \bibinfo
  {author} {\bibfnamefont {B}~\bibnamefont {Dunham}},\ }\bibfield  {title}
  {\bibinfo {title} {Cold electron beams from cryocooled, alkali antimonide
  photocathodes},\ }\href@noop {} {\bibfield  {journal} {\bibinfo  {journal}
  {Physical Review Special Topics-Accelerators and Beams}\ }\textbf {\bibinfo
  {volume} {18}},\ \bibinfo {pages} {113401} (\bibinfo {year}
  {2015})}\BibitemShut {NoStop}%
\bibitem [{\citenamefont {Musumeci}\ \emph {et~al.}(2018)\citenamefont
  {Musumeci}, \citenamefont {Navarro}, \citenamefont {Rosenzweig},
  \citenamefont {Cultrera}, \citenamefont {Bazarov}, \citenamefont {Maxson},
  \citenamefont {Karkare},\ and\ \citenamefont
  {Padmore}}]{musumeci2018advances}%
  \BibitemOpen
  \bibfield  {author} {\bibinfo {author} {\bibfnamefont {P}~\bibnamefont
  {Musumeci}}, \bibinfo {author} {\bibfnamefont {J~Giner}\ \bibnamefont
  {Navarro}}, \bibinfo {author} {\bibfnamefont {JB}~\bibnamefont {Rosenzweig}},
  \bibinfo {author} {\bibfnamefont {L}~\bibnamefont {Cultrera}}, \bibinfo
  {author} {\bibfnamefont {I}~\bibnamefont {Bazarov}}, \bibinfo {author}
  {\bibfnamefont {J}~\bibnamefont {Maxson}}, \bibinfo {author} {\bibfnamefont
  {S}~\bibnamefont {Karkare}}, and\ \bibinfo {author} {\bibfnamefont
  {H}~\bibnamefont {Padmore}},\ }\bibfield  {title} {\bibinfo {title} {Advances
  in bright electron sources},\ }\href@noop {} {\bibfield  {journal} {\bibinfo
  {journal} {Nuclear Instruments and Methods in Physics Research Section A:
  Accelerators, Spectrometers, Detectors and Associated Equipment}\ }\textbf
  {\bibinfo {volume} {907}},\ \bibinfo {pages} {209--220} (\bibinfo {year}
  {2018})}\BibitemShut {NoStop}%
\bibitem [{\citenamefont {Franssen}\ and\ \citenamefont
  {Luiten}(2017)}]{Franssen2017}%
  \BibitemOpen
  \bibfield  {author} {\bibinfo {author} {\bibfnamefont {JGH}\ \bibnamefont
  {Franssen}}and\ \bibinfo {author} {\bibfnamefont {OJ}~\bibnamefont
  {Luiten}},\ }\bibfield  {title} {\bibinfo {title} {Improving temporal
  resolution of ultrafast electron diffraction by eliminating arrival time
  jitter induced by radiofrequency bunch compression cavities},\ }\href@noop {}
  {\bibfield  {journal} {\bibinfo  {journal} {Structural Dynamics}\ }\textbf
  {\bibinfo {volume} {4}},\ \bibinfo {pages} {044026} (\bibinfo {year}
  {2017})}\BibitemShut {NoStop}%
\bibitem [{\citenamefont {van~der Geer}\ and\ \citenamefont
  {De~Loos}(2001)}]{van2001general}%
  \BibitemOpen
  \bibfield  {author} {\bibinfo {author} {\bibfnamefont {Bas}\ \bibnamefont
  {van~der Geer}}and\ \bibinfo {author} {\bibfnamefont {MJ}~\bibnamefont
  {De~Loos}},\ }\href@noop {} {\emph {\bibinfo {title} {The general particle
  tracer code: design, implementation and application}}}\ (\bibinfo
  {publisher} {Technische Universiteit Eindhoven},\ \bibinfo {year}
  {2001})\BibitemShut {NoStop}%
\bibitem [{\citenamefont {Karkare}\ \emph {et~al.}(2020)\citenamefont
  {Karkare}, \citenamefont {Adhikari}, \citenamefont {Schroeder}, \citenamefont
  {Nangoi}, \citenamefont {Arias}, \citenamefont {Maxson},\ and\ \citenamefont
  {Padmore}}]{karkare2020ultracold}%
  \BibitemOpen
  \bibfield  {author} {\bibinfo {author} {\bibfnamefont {Siddharth}\
  \bibnamefont {Karkare}}, \bibinfo {author} {\bibfnamefont {Gowri}\
  \bibnamefont {Adhikari}}, \bibinfo {author} {\bibfnamefont {W~Andreas}\
  \bibnamefont {Schroeder}}, \bibinfo {author} {\bibfnamefont {J~Kevin}\
  \bibnamefont {Nangoi}}, \bibinfo {author} {\bibfnamefont {Tomas}\
  \bibnamefont {Arias}}, \bibinfo {author} {\bibfnamefont {Jared}\ \bibnamefont
  {Maxson}}, and\ \bibinfo {author} {\bibfnamefont {Howard}\ \bibnamefont
  {Padmore}},\ }\bibfield  {title} {\bibinfo {title} {Ultracold electrons via
  near-threshold photoemission from single-crystal cu (100)},\ }\href@noop {}
  {\bibfield  {journal} {\bibinfo  {journal} {arXiv preprint arXiv:2002.11579}\
  } (\bibinfo {year} {2020})}\BibitemShut {NoStop}%
\bibitem [{\citenamefont {Baum}(2017)}]{baum2017quantum}%
  \BibitemOpen
  \bibfield  {author} {\bibinfo {author} {\bibfnamefont {Peter}\ \bibnamefont
  {Baum}},\ }\bibfield  {title} {\bibinfo {title} {Quantum dynamics of
  attosecond electron pulse compression},\ }\href@noop {} {\bibfield  {journal}
  {\bibinfo  {journal} {Journal of Applied Physics}\ }\textbf {\bibinfo
  {volume} {122}},\ \bibinfo {pages} {223105} (\bibinfo {year}
  {2017})}\BibitemShut {NoStop}%
\bibitem [{\citenamefont {Cardin}\ \emph {et~al.}(2018)\citenamefont {Cardin},
  \citenamefont {Jeong}, \citenamefont {Fan}, \citenamefont {Balciunas},
  \citenamefont {Piccoli}, \citenamefont {Ferachou}, \citenamefont {Limpert},
  \citenamefont {H{\"a}drich}, \citenamefont {Morandotti}, \citenamefont
  {W{\"o}rner} \emph {et~al.}}]{cardin2018high}%
  \BibitemOpen
  \bibfield  {author} {\bibinfo {author} {\bibfnamefont {Vincent}\ \bibnamefont
  {Cardin}}, \bibinfo {author} {\bibfnamefont {Young-Gyun}\ \bibnamefont
  {Jeong}}, \bibinfo {author} {\bibfnamefont {Guangyu}\ \bibnamefont {Fan}},
  \bibinfo {author} {\bibfnamefont {Tadas}\ \bibnamefont {Balciunas}}, \bibinfo
  {author} {\bibfnamefont {Riccardo}\ \bibnamefont {Piccoli}}, \bibinfo
  {author} {\bibfnamefont {Denis}\ \bibnamefont {Ferachou}}, \bibinfo {author}
  {\bibfnamefont {Jens}\ \bibnamefont {Limpert}}, \bibinfo {author}
  {\bibfnamefont {Steffen}\ \bibnamefont {H{\"a}drich}}, \bibinfo {author}
  {\bibfnamefont {Roberto}\ \bibnamefont {Morandotti}}, \bibinfo {author}
  {\bibfnamefont {Hans~Jakob}\ \bibnamefont {W{\"o}rner}},  \emph {et~al.},\
  }\bibfield  {title} {\bibinfo {title} {High power hollow-core fiber
  compression of yb lasers as ideal drivers for hhg},\ }in\ \href@noop {}
  {\emph {\bibinfo {booktitle} {CLEO: Science and Innovations}}}\ (\bibinfo
  {organization} {Optical Society of America},\ \bibinfo {year} {2018})\ pp.\
  \bibinfo {pages} {SM4M--3}\BibitemShut {NoStop}%
\bibitem [{\citenamefont {Zuo}\ and\ \citenamefont
  {Spence}(2017)}]{zuo2017advanced}%
  \BibitemOpen
  \bibfield  {author} {\bibinfo {author} {\bibfnamefont {Jian~Min}\
  \bibnamefont {Zuo}}and\ \bibinfo {author} {\bibfnamefont {John~CH}\
  \bibnamefont {Spence}},\ }\bibfield  {title} {\bibinfo {title} {Advanced
  transmission electron microscopy},\ }\href@noop {} {\bibfield  {journal}
  {\bibinfo  {journal} {Advanced Transmission Electron Microscopy, ISBN
  978-1-4939-6605-9. Springer Science+ Business Media New York, 2017}\ }
  (\bibinfo {year} {2017})}\BibitemShut {NoStop}%
\end{thebibliography}
\end{document}